\documentclass[12pt,a4paper,notitlepage]{article}

\usepackage{amsfonts}
\usepackage{amsmath}
\usepackage{amssymb}
\usepackage{amsthm}
\usepackage{color}
\usepackage[T1]{fontenc}
\usepackage{graphicx}
\usepackage{hyperref}
\usepackage{mathrsfs}
\usepackage{multirow}
\usepackage[numbers]{natbib}
\usepackage[format=hang]{subcaption}
\usepackage{times}
\usepackage{upgreek}
\usepackage[utf8]{inputenc}
\usepackage{pgfplots}
\usepackage{longtable}
\usepackage{booktabs}
\usepackage{algorithm, algpseudocode}
\usepackage{adjustbox}
\usepackage{siunitx}
\usepackage{svg}
\usepackage{tikz}
\usepackage{import}
\usetikzlibrary{patterns,shapes.arrows}
\pgfplotsset{compat=newest}
\usepackage{float}
\pgfplotsset{compat=newest}
\usetikzlibrary{plotmarks}
\usetikzlibrary{arrows.meta}
\usetikzlibrary{calc}
\usepackage{grffile}
\usepackage{array, makecell} 
\usepackage{tabularx}
\AtBeginDocument{}

\date{}

\graphicspath{{./images/}}

\newcounter{remark}

\begin{document}
\author{ {M. Sesa${}^{\,\mathrm{a}}$, H. Holthusen${}^{\,\mathrm{a}}$, C. B\"ohm${}^{\,\mathrm{b}}$,} \\ { S. Jockenh\"ovel${}^{\,\mathrm{b}}$, S. Reese${}^{\,\mathrm{a, \, c}}$, K. Linka${}^{\,\mathrm{a, \, d}}$}\\[0.5cm]
\normalsize{\em ${}^{\mathrm{a}}$Institute of Applied Mechanics, RWTH Aachen
  University, Germany}\\\
\normalsize{\em ${}^{\mathrm{b}}$ Department of Biohybrid \& Medical Textiles (BioTex), Institute of Applied Medical Engineering, } \\ [-0.1cm] \normalsize{\em RWTH Aachen University, Germany}\\
\normalsize{\em ${}^{\mathrm{c}}$ University of Siegen, Germany}\\
\normalsize{\em ${}^{\mathrm{d}}$ Institute for Continuum and Material Mechanics, Hamburg University of Technology, Germany}\\
}
\title{\LARGE A Comprehensive Framework for Predictive Computational Modeling of Growth and Remodeling in Tissue-Engineered Cardiovascular Implants}
\maketitle

\small
{\bf Abstract.}
Developing clinically viable tissue-engineered cardiovascular implants remains a formidable challenge. Achieving reliable and durable outcomes requires a deeper understanding of the fundamental mechanisms driving tissue evolution during in vitro maturation. Although considerable progress has been made in modeling soft tissue growth and remodeling, studies focused on the early stages of tissue engineering remain limited. Here, we present a general, thermodynamically consistent model to predict tissue evolution and mechanical response throughout maturation. The formulation utilizes a stress-driven homeostatic surface to capture volumetric growth, coupled with an energy-based approach to describe collagen densification via the strain energy of the fibers. We further employ a co-rotated intermediate configuration to ensure the model's consistency and generality. The framework is demonstrated with two numerical examples: a uniaxially constrained tissue strip validated against experimental data, and a biaxially constrained specimen subjected to a perturbation load. These results highlight the potential of the proposed model to advance the design and optimization of tissue-engineered implants with clinically relevant performance.

\vspace*{0.3cm}

\newpage

\normalsize


\section{Introduction}
\label{sec:1}
Cardiovascular diseases (CVDs) are the primary cause of death worldwide \cite{Di_Cesare_2024}. Deaths attributed to CVDs have sharply increased in the last few decades \cite{Amini_2021}. Tissue-engineered implants offer a promising long-term solution to treat CVDs by improving patients' lifestyles compared to alternative implant types. This stems from their ability to grow, remodel, and adapt to hemodynamic conditions \cite{Yacoub_Takkenberg_2005, Pashneh_2016}, combined with their resistance to calcification \cite{Turner_2024}, making them a viable option to alleviate the limitations of existing cardiovascular implants. Tissue-engineered implants are fabricated in a process called the maturation process. Developing a maturation process capable of producing complex cardiovascular implants, such as heart valves or vascular grafts with sufficient mechanical properties remains a challenging task.

The maturation process is influenced by numerous factors, such as the cultivation medium composition, the applied loading and boundary conditions, and the process duration. Another major factor is whether the tissue is reinforced or not, as well as the reinforcement type and properties. All these factors affect the cultivation process, yet their influence remains not well understood. Additional complications arise from the difficulty of experimentally replicating the conditions within the human body, leading to discrepancies between the results of in vitro and in vivo experiments. An example is the concept of contact guidance which is used to tailor the mechanical properties of textile-based tissue-engineered implants. Contact guidance was found to be ineffective in large animal experiments \cite{Uiterwijk_2020},  contradicting findings from in vitro experiments \cite{Foolen_2012, Hermans_etal_2022}. This underscores the necessity to develop a fundamental understanding of tissue-engineered material's mechanobiology. The regulatory and financial limitations on performing large animal experiments demonstrate the need for accurate predictive in silico models. Furthermore, in silicon models together with experiments provide deeper insights into the growth and remodeling process, which are necessary to optimize the maturation process and the implant design.

The maturation process involves the secretion of extracellular matrix (ECM) which provides structural support for the tissue through a network of protein fibers such as elastin and collagen fibers. Collagen fibers are the primary structural constituent of the tissue. The changes in the mechanical properties of the tissue are driven by the evolution of collagen density and orientation. Therefore, investigating the mechanisms controlling the evolution of collagen fibers has been widely investigated \cite{Huang_Yannas_1977, Wyatt_etal_2009, Siadat_Ruberti_2023, Ruberti_Hallab_2005}. During the maturation process, internal stresses develop within the biological tissue. Biological tissues achieve equilibrium at a certain level of stress, called homeostatic stress, in a phenomenon known as tensional homeostasis \cite{Stamenovic_2020}. Experimental investigations by Eichinger et al.\ \cite{Eichinger_2021} demonstrated that the homeostatic stress level depends on collagen density. The process of regulating tissue homeostatic stress drives changes in tissue shape and the reorientation of collagen fibers. ECM homeostasis has been thoroughly investigated in the context of cancer research. Investigations by Paszek et al.\ \cite{Paszek_2005} revealed that the stiffness of the ECM influences cell contractility and the tensional homeostatic behavior of the tissue. These findings align with the results obtained from studies on soft tissue constructs by Eichinger et al.\ \cite{Eichinger_2021}. In another study, Cox \& Erler \cite{Cox_2011} investigated the influence of homeostatic conditions on ECM remodeling. 

Researchers have explored various approaches to model biological growth and remodeling. Common approaches are (i) kinematic-based models \cite{Rodriguez_1994}, (ii) constrained mixture models \cite{Humphrey_2002}, and (iii) agent-based models \cite{Bonabeau_2002, An_2009}. Kinematic models are particularly advantageous due to their computational efficiency. Constrained mixture models require defining material parameters for each tissue constituent such as smooth muscle cells, elastin, and collagen fibers. Experimental identification of these parameters is highly challenging. In this work, we propose a kinematic-based framework. Biological growth occurs either as changes in shape, which is referred to as volumetric growth, or changes in densities of tissue constituents. Both forms alter the total mass of the system. Kinematic models describe volumetric growth using the multiplicative split of the deformation gradient into elastic and inelastic growth parts \cite{Rodriguez_1994}. The growth part is called the growth tensor. This approach has been applied to various problems, including studying growth mechanisms in heart valves \cite{Oomen_etal_2018} and in-stent restenosis \cite{Manjunatha_2022, Manjunatha_2023}. Reviews on growth models and their applications can be found in \cite{Ambrosi_2011, Kuhl_2014, Eskandari_2015, Ambrosi_2019}. Initial studies relied on heuristic assumptions about shape evolution to predefine the orientation of the growth tensor. Although this approach is simple to implement, it limits the predictive capabilities of the model to solve general boundary value problems, leading to unphysical results in constrained growth problems \cite{Braeu_2017, Braeu_2019, Soleimani_2020, Lamm_2022}. To address these limitations, Braeu et al.\ \cite{Braeu_2019} developed an anisotropic growth model based on the concept of homeostatic stress, while Soleimani et al.\ \cite{Soleimani_2020} introduced a stress-driven anisotropic growth framework. Another approach was proposed in Lamm et al.\ \cite{Lamm_2022} which derives growth tensors from the homeostatic stress surface defined in the principal stress space. This ensures thermodynamic consistency and accurately describes the evolution of residual stresses. Later Holthusen et al.\ \cite{Holthusen_2023} developed a two-surface model for growth and remodeling. Additionally, the physics-based inelastic Constitutive Artificial Neural Networks framework \cite{Holthusen_2024b} has demonstrated promising results in modeling volumetric growth.

Despite extensive research in biomechanics on modeling growth and remodeling, most models focus on native tissues. Studies addressing modeling the tissue-engineering process are limited. Szafron et al.\ \cite{Szafron_etal_2019} demonstrated that numerical models can significantly improve the design of tissue-engineered implants. However, the model is only valid for a specific experimental setup. Similarly, Loerakker et al.\  \cite{Loerakker_2013,Loerakker_2016} and Sanders et al.\ \cite{Sanders_2016} provided valuable insights into heart valve mechanobiology. However, these models neglected volumetric growth and lacked proof of satisfying the laws of thermodynamics. The constitutive model introduced in Sesa et al.\ \cite{Sesa_2023_CBM} to model the maturation process of textile-reinforced tissue-engineered implants considers collagen density evolution to be driven by biochemical and mechanobiological factors. The strain energy in collagen fibers was chosen as the driving factor for the mechanobiological stimulation, ensuring the thermodynamic consistency of the model. However, the model neglected volumetric growth and fiber reorientation. Consequently, the cumulative increase in internal stresses driven by shape evolution was neglected. A more realistic model should consider volumetric growth, as well as the evolution of collagen density and orientation. Modeling these interdependent phenomena requires constructing a coupled system of evolution equations to fully capture the behavior of the tissue. The models for collagen densification \cite{Sesa_2023_CBM}, as well as volumetric growth and remodeling \cite{Holthusen_2023} provide a solid theoretical basis for the construction of a generalized and thermodynamically consistent growth model for tissue-engineered implants.

In this contribution, we introduce a predictive model that describes the evolution of biological tissue mechanical behavior during the maturation process. The model describes the underlying phenomena of collagen fiber build-up and the shape evolution. During the maturation process, biological tissues undergo significant changes in their mass, volume, and mechanical properties. This unique behavior renders models developed to describe the evolution of native tissues unsuitable. Existing models developed to optimize the implant's design in a specific setup \cite{Szafron_etal_2019}, are valuable. However, without developing objective and thermodynamically consistent constitutive models, it is doubtful that we can build accurate predictive models for tissue engineering processes. This motivated us to develop a general and thermodynamically consistent approach. 

This work presents significant distinctions from previous studies. The study presented in \cite{Sesa_2023_CBM} focused on modeling collagen density evolution during the maturation of textile-reinforced implants while neglecting collagen reorientation and volumetric growth. In this contribution, the model was extended by incorporating volumetric growth and collagen reorientation to model unreinforced tissues. This distinguishes our approach from \cite{Holthusen_2023}, where collagen density evolution is neglected. Another distinction is our focus here on modeling tissues that initially lack any collagen content and computing their evolution for a period of four weeks, compared to 1-2 days in \cite{Holthusen_2023}. These considerations require for the first time defining the homeostatic stress as a function of collagen density, rather than a constant \cite{Lamm_2022, Holthusen_2023}. Additionally, we introduce a one-surface approach with two pseudo-potentials, which combine the benefits of a two-surface approach in separately describing the evolution of the matrix and collagen parts \cite{Holthusen_2023}, while eliminating the need to define two homeostatic stress parameters for the matrix and collagen parts, which are experimentally challenging to identify. 

The next sections are organized as follows. Section \ref{sec:2}, introduces the theoretical aspects of the model, including constitutive laws and evolution equations. Then, section \ref{sec:3} provides a concise description of the finite element implementation. In section \ref{sec:4}, we compute two structural examples.  The first example mimics the experimental setup for the in vitro maturation of uniaxially constrained construct. Experimental data are used to identify material parameters and validate the numerical model. In the second example, we computed the evolution of a tissue-engineered construct subjected to biaxial loading conditions and load perturbation. The aim is to study the capabilities of our framework in describing structures under complex loading conditions. Finally, in section \ref{sec:5}, we present our conclusion and outlook for future studies.

\section{Continuum mechanics model}
\label{sec:2}
This section presents a constitutive modeling framework based on continuum mechanics to describe the evolution of biological material during maturation. To ensure the simplicity of the model, we focus only on constituents and mechanisms that significantly affect the material's mechanical response. A common approach to model soft collagenous tissues is to split the total Helmholtz free energy into an isotropic matrix part and an anisotropic collagen part. The validity of this modeling approach for tissue-engineered constructs was confirmed in \cite{Sesa_2023_CBM}, which measured the stress-strain response of biological constructs and the corresponding alterations in collagen density at different time points during the maturation process. Building on this work, we propose here a model that better describes the evolution of the tissue's mechanical properties during the maturation process. The model considers three phenomena as the main factors influencing the mechanical response: i) collagen density distribution, ii) collagen orientation, and iii) volumetric growth.

We start by introducing the kinematic relations and the relevant balance equations. Then, we derive the Clausius-Duhem inequality. Next, we extend our equations to account for the decomposition of the Helmholtz free energy into an isotropic matrix and anisotropic collagen parts. Afterward, we introduce the evolution equations describing volumetric growth, collagen density evolution, and fiber reorientation. Finally, we define specific Helmholtz free energy functions chosen for the matrix and collagen parts.

\subsection{Kinematics}
The first step in constructing our continuum mechanics model is to define the kinematic relations. In a three-dimensional continuum body, the function $\phi$ maps between the reference configuration $\Omega_{0}$ at time $t = 0$ and the current configuration $\Omega$ at time $t$. The position vector for a material particle in the reference configuration is $\mathbf{X}$, and the corresponding position vector in the current configuration is $\mathbf{x} = \phi(\mathbf{X}, t) $. The mapping from the reference configuration to the current configuration is achieved by applying the deformation gradient tensor $\mathbf{F} = \partial \mathbf{x}/ \partial \mathbf{X}$. Then we obtain the right Cauchy-Green tensor $ \mathbf{C} = \mathbf{F}^{\mathrm{T}} \, \mathbf{F} $ and the left Cauchy-Green tensor $\mathbf{B} = \mathbf{F} \, \mathbf{F}^{\mathrm{T}}$. 

The growth of biological materials is an inelastic process. The kinematics of such a process can be described using the multiplicative decomposition of the deformation gradient 
\begin{equation}
\label{F_multi_split}
\mathbf{F} = \mathbf{F}_{\mathrm{e}} \, \mathbf{F}_{\mathrm{g}},
\end{equation}
where $\mathbf{F}_{\mathrm{g}}$ is the growth part, and $\mathbf{F}_{\mathrm{e}}$ is the elastic part \cite{Rodriguez_1994}. In the next step, we define the elastic and growth-related right Cauchy-Green tensors respectively:
\begin{equation}
\begin{split}
\label{}
\mathbf{C}_{\mathrm{e}} &:= \mathbf{F}^{\mathrm{T}}_{\mathrm{e}} \, \mathbf{F}_{\mathrm{e}} =  \mathbf{F}^{\mathrm{-T}}_{\mathrm{g}} \, \mathbf{C} \, \mathbf{F}_{\mathrm{e}}, \\
\mathbf{C}_{\mathrm{g}} &:= \mathbf{F}^{\mathrm{T}}_{\mathrm{g}} \, \mathbf{F}_{\mathrm{g}}. 
\end{split}
\end{equation}

The tensor $\mathbf{F}_{\mathrm{g}}$ maps the reference configuration $\Omega_{0}$ to a stress-free intermediate configuration. This intermediate configuration is characterized by its rotational non-uniqueness. The polar decomposition operation $\mathbf{F}_{\mathrm{g}} = \mathbf{R}_{\mathrm{g}} \, \mathbf{U}_{\mathrm{g}}$ splits the tensor into the rotation tensor $\mathbf{R}_{\mathrm{g}}$ and the right stretch tensor $ \mathbf{U}_{\mathrm{g}} $. From this split, it was identified that $ \mathbf{U}_{\mathrm{g}} $ is uniquely defined, while  $\mathbf{R}_{\mathrm{g}}$ suffers from rotational non-uniqueness. To address this issue, we apply the Holthusen et al.\ framework \cite{Holthusen_2023} which proposed performing a pull-back operation of the kinematic quantities and structural tensors to a uniquely defined configuration called the co-rotated intermediate configuration (\textit{cic}). By applying this operation, the right Cauchy-Green tensor in \textit{cic} becomes 
\begin{equation}
\label{C_cic}
\bar{\mathbf{C}}_{\mathrm{e}} := \mathbf{R}^{-1}_{\mathrm{g}} \, \mathbf{C}_{\mathrm{e}} \, \mathbf{R}_{\mathrm{g}} = \mathbf{U}^{-1}_{\mathrm{g}} \, \mathbf{C} \, \mathbf{U}^{-1}_{\mathrm{g}},
\end{equation}
where we refer to quantities defined in \textit{cic} using the notation $(\, \bar{} \,)$.

\subsection{Balance of linear momentum}
When modeling the maturation process of biological tissues, various time scales must be considered. The growth and remodeling processes take place over weeks, while the elastic response occurs on the order of milliseconds. This means that our model can be simplified by applying the slow-growth assumption \cite{Goriely_2017}, which means that the balance of mass is satisfied without additional considerations, as demonstrated in \cite{Sesa_2023_CBM}. Consequently, the inertia effect from the added mass is negligibly small, allowing us to consider our system to be quasi-static, and apply the standard balance of linear momentum equation  
\begin{equation}
\label{balance_linear_momentum}
\mathrm{Div}({\mathbf F} \, {\mathbf S}) + {\mathbf b}_{0} = {\mathbf 0},
\end{equation}
where ${\mathbf S}$ is the second Piola-Kirchhoff stress, and ${\mathbf b}_{0}$ is the body force vector in the reference configuration.

\subsection{Clausius-Duhem inequality}
\label{subsec:2-2}
The Clausius-Duhem inequality is
\begin{equation}
\label{Clausius-Duhem}
\frac{1}{2} \, {\mathbf S} \cdot \dot{{\mathbf C}} - \dot{\psi} + S_{\mathrm{0}} \geqslant 0,
\end{equation}
where $S_{0}$ is an additional term that accounts for both the local entropy production and the entropy flux through the boundaries of an open system \cite{Kuhl_Steinmann_2003}. The material time derivative is denoted by the shorthand notation $({\bullet})$.

During the maturation process, the build up of collagen content significantly influence the mechanical response of the material. This necessitates introducing the collagen density in the reference configuration ${\rho}^{0}_{\mathrm{co}}$ as an argument in the Helmholtz free energy function ${\psi}$ \cite{Sesa_2023_CBM} as expressed here 
\begin{equation}
	\psi = \bar{\psi}(\bar{\mathbf{C}}_{\mathrm{e}} , \bar{\mathbf{H}}, {\rho}^{0}_{\mathrm{co}}).
\end{equation}
 
The material rate of the Helmholtz free energy becomes
\begin{equation}
\dot{\psi} = \dot{\bar{\psi}} = \frac{\partial \bar{\psi}}{\partial \bar{\mathbf{C}}_{\mathrm{e}}} \cdot \dot{\bar{\mathbf{C}}}_{\mathrm{e}} \, + \, \frac{\partial \bar{\psi}}{\partial \bar{\mathbf{H}}} \cdot \dot{\bar{\mathbf{H}}} \, + \, \frac{\partial \bar{\psi}}{{\rho}^{0}_{\mathrm{co}}} \dot{{{\rho}}}^{0}_{\mathrm{co}},
\end{equation}
where $\bar{\mathbf{C}}_{\mathrm{e}}$, $\bar{\mathbf{H}}$ and are defined in the \textit{cic}, and ${\rho}^{0}_{\mathrm{co}}$ is defined in the reference configuration.

To reformulate the Clausius-Duhem inequality in Eq.\ (\ref{Clausius-Duhem}), we introduce the growth-related velocity gradient in the \textit{cic}
\begin{equation}
\label{velocity_gradient}
\bar{\mathbf{L}}_{\mathrm{g}} = \dot{\mathbf{U}}_{\mathrm{g}} \, \mathbf{U}^{-1}_{\mathrm{g}}.
\end{equation}

Using the definition of  $\bar{\mathbf{L}}_{\mathrm{g}}$ in Eq.\ (\ref{velocity_gradient}), we can formulate the following rate quantities
\begin{equation}
\dot{\bar{\mathbf{C}}}_{\mathrm{e}}  = {\mathbf{U}}^{-1}_{\mathrm{g}} \, \dot{\mathbf{C}} \, \mathbf{U}^{-1}_{\mathrm{g}} - \bar{\mathbf{C}}_{\mathrm{e}} \, \bar{\mathbf{L}}_{\mathrm{g}} - \bar{\mathbf{L}}^{T}_{\mathrm{g}} \, \bar{\mathbf{C}}_{\mathrm{e}},
\end{equation}
\begin{equation}
\dot{\bar{\mathbf{H}}} = \mathbf{U}_{\mathrm{g}} \, \dot{\mathbf{H}} \, \mathbf{U}_{\mathrm{g}} + \bar{\mathbf{L}}_{\mathrm{g}} \, \bar{\mathbf{H}} + \bar{\mathbf{H}} \, \bar{\mathbf{L}}^{T}_{\mathrm{g}}.
\end{equation}

From the Coleman-Noll procedure \cite{Coleman_Noll_1963}, we get the following expression for the second Piola-Kirchhoff stress 
\begin{equation}
\label{2nd_Piola_Kirchhoff}
\mathbf{S} = 2 \, \mathbf{U}^{-1}_{\mathrm{g}}  \, \frac{\partial \bar{\psi}}{\partial \bar{\mathbf{C}}_{\mathrm{e}}} \, \mathbf{U}^{-1}_{\mathrm{g}},
\end{equation}
and the reduced dissipation inequality becomes
\begin{equation}
\label{dissipation_inequality}
\mathscr{D}_{\mathrm{red}} := \underbrace{ \left(\overbrace{2 \, \bar{\mathbf{C}}_{\mathrm{e}} \, \frac{\partial \bar{\psi}}{\partial \bar{\mathbf{C}}_{\mathrm{e}}}}^{=: \, \bar{\mathbf{\Sigma}}}  \, - \, \overbrace{2 \, \frac{\partial \bar{\psi}}{\partial \bar{\mathbf{H}}} \, \bar{\mathbf{H}}}^{=: \, \bar{\mathbf{Y}}}  \right)}_{\bar{\mathbf{\Gamma}}} \cdot \bar{\mathbf{L}}_{\mathrm{g}} \, - \, \underbrace{ \mathbf{U}_{\mathrm{g}} \, \frac{\partial \bar{\psi}}{\partial \bar{\mathbf{H}}}  \, \mathbf{U}_{\mathrm{g}} }_{=: \,  {\mathbf{G}}} \cdot \dot{\mathbf{H}} \, - \, \frac{\partial \bar{\psi}}{{\partial \rho^{0}}_{\mathrm{co}}} \, \dot{{\rho}}^{0}_{\mathrm{co}} \, + \, S_{0} \, \geq 0,
\end{equation}
where $\bar{\mathbf{\Sigma}}$ and $\bar{\mathbf{Y}}$ are stress-like tensor quantities defined in the \textit{cic}.

By exploiting the symmetry of the tensor $\bar{\mathbf{\Gamma}}$ \cite{Svendsen_2001, Reese_2003, Holthusen_2023}, the dissipation inequality in Eq.\ (\ref{dissipation_inequality}) can be rewritten as
\begin{equation}
   \label{dissipation_inequality_1}
	\mathscr{D}_{\mathrm{red}} := \bar{\mathbf{\Gamma}} \cdot \bar{\mathbf{D}}_{\mathrm{g}} \, - \, {\mathbf{G}} \cdot \dot{\mathbf{H}} \, - \, \frac{\partial \bar{\psi}}{{\partial \rho^{0}}_{\mathrm{co}}} \, \dot{{\rho}}^{0}_{\mathrm{co}} \, + \, S_{0} \geq 0,
\end{equation}
with $\bar{\mathbf{D}}_{\mathrm{g}}  := \mathrm{sym}(\bar{\mathbf{L}}_{\mathrm{g}})$.

\subsection{Extension to a multi-constituent material}
\label{subsec:2-3}
The ECM in tissue-engineered materials consists of various constituents, with fibrous constituents such as collagen and elastin fibers significantly affecting the tissue's mechanical behavior. Microscopy images reveal that the elastin fibers are short and dispersed, while the collagen fibers are long, rigid, and highly oriented fibers. Thus, decomposing the material into an isotropic ground matrix and anisotropic collagen fibers is applicable here as demonstrated in Sesa et al.\ \cite{Sesa_2023_CBM}. Based on this, we formulate the total Helmholtz free energy as
\begin{equation}
\psi = \bar{\psi}(\bar{\mathbf{C}}_{\mathrm{e}} , \bar{\mathbf{H}}, {\rho}^{0}_{\mathrm{co}}) \, = \bar{\psi}_{\mathrm{m}}(\bar{\mathbf{C}}_{\mathrm{e}_{\mathrm{m}}} ) \, + \, \bar{\psi}_{\mathrm{co}}(\bar{\mathbf{C}}_{\mathrm{e}_{\mathrm{co}}}, \bar{\mathbf{H}}, {\rho}^{0}_{\mathrm{co}}) ,
\end{equation}
where ${\psi}_{\mathrm{m}}$ and ${\psi}_{\mathrm{co}}$ refer to the Helmholtz free energies for the matrix and collagen parts, respectively. In a similar manner, we decompose the deformation gradient into a matrix part (m) and a collagen part (co) \cite{Holthusen_2023}
\begin{equation}
\label{F_co_m}
\mathbf{F} := \mathbf{F}_{\mathrm{e}_{\mathrm{m}}} \, \mathbf{F}_{\mathrm{g}_{\mathrm{m}}} := \mathbf{F}_{\mathrm{e}_{\mathrm{co}}} \, \mathbf{F}_{\mathrm{g}_{\mathrm{co}}}.
\end{equation}

From the deformation gradients introduced in Eq.\ (\ref{F_co_m}), we get the corresponding elastic part of the right Cauchy-Green tensor
\begin{equation}
\label{}
 \mathbf{C}_{\mathrm{e}_{\mathrm{j}}} := \mathbf{F^{T}}_{\mathrm{e}_{\mathrm{j}}} \, \mathbf{F}_{\mathrm{e}_{\mathrm{j}}},
\end{equation}
where we refer to the matrix and collagen parts using the index $\mathrm{j}=\{\mathrm{m}, \mathrm{co} \}$. This is then transformed to the \textit{cic} by applying the same procedure presented in Eq.\ (\ref{C_cic}), to get
\begin{equation}
\bar{\mathbf{C}}_{\mathrm{e}_{\mathrm{j}}} := \mathbf{U}^{-1}_{\mathrm{g}_{\mathrm{j}}} \, \mathbf{C} \, \mathbf{U}^{-1}_{\mathrm{g}_{\mathrm{j}}}.
\end{equation}

The next step is to introduce the anisotropic behavior of collagen fibers into the constitutive equations using structural tensors. For a vector $\mathbf{a}$ which defines the mean orientation of the collagen fibers, we get the following structural tensor
\begin{equation}
\label{structural_tensor} 
\mathbf{M} := {\mathbf a} \otimes {\mathbf a}.
\end{equation}
Then can be transformed to the \textit{cic} using the transformation process used in \cite{Holthusen_2023, Reese_2003}
\begin{equation}
\bar{\mathbf{M}} = \frac{1}{\mathbf{C}_{\mathrm{g}_{\mathrm{co}}}:\mathbf{M}} \, \mathbf{U}_{\mathrm{g}_{\mathrm{co}}} \, \mathbf{M} \, \mathbf{U}_{\mathrm{g}_{\mathrm{co}}}.
\end{equation}

Furthermore, it is important to consider the dispersion of fiber orientation in collagen bundles. Therefore, we apply the generalized structural tensor formulation from Gasser et al.\ \cite{Gasser_etal_2006}
\begin{equation}
\bar{\mathbf{H}} := \kappa \, {\mathbf{I}} \, + \, (1 \, - \, 3 \, \kappa) \, \bar{\mathbf{M}},
\end{equation}
where $\bar{\mathbf{H}}$ lies in the \textit{cic}, and $ 0 \le {\kappa} \le \frac{1}{3}$. 

By applying the decomposition into the matrix and collagen parts to the expression of the second Piola-Kirchhoff stress in Eq.\ (\ref{2nd_Piola_Kirchhoff}), we obtain the following expression 
\begin{equation}
\mathbf{S} = \sum_{\mathrm{j}} \mathbf{S}_{\mathrm{j}} = 2 \, \sum_{\mathrm{j}}  \mathbf{U}^{-1}_{\mathrm{g}_{\mathrm{j}}}  \, \frac{\partial \bar{\psi}}{\partial \bar{\mathbf{C}}_{\mathrm{e}_{\mathrm{j}}}} \, \mathbf{U}^{-1}_{\mathrm{g}_{\mathrm{j}}}.
\end{equation} 

Furthermore, the reduced dissipation inequality in Eq.\ (\ref{dissipation_inequality_1}) can be rewritten as
\begin{equation}
\label{dissipation_inequality_2}
\mathscr{D}_{\mathrm{red}} := \underbrace{ \bar{\mathbf{\Sigma}}_{\mathrm{m}} }_{=: \,  \bar{\mathbf{\Gamma}}_{\mathrm{m}}}  \cdot \bar{\mathbf{L}}_{\mathrm{g}_{\mathrm{m}}} + \underbrace{ \left( \bar{\mathbf{\Sigma}}_{\mathrm{co}} -  \bar{\mathbf{Y}}_{\mathrm{co}} + \bar{\mathbf{\Pi}}_{\mathrm{co}} \right)  }_{=: \,  \bar{\mathbf{\Gamma}}_{\mathrm{co}}} \cdot \bar{\mathbf{L}}_{\mathrm{g}_{\mathrm{co}}} \, - \, \mathbf{G}_{\mathrm{co}} \cdot \dot{\mathbf{H}} \, - \, \frac{\partial \bar{\psi}}{{\partial \rho^{0}}_{\mathrm{co}}} \, \dot{{\rho}}^{0}_{\mathrm{co}} \, + \, S_{0} \, \geq 0,
\end{equation}
with $\bar{\mathbf{\Sigma}}_{\mathrm{m}} := 2 \, \bar{\mathbf{C}}_{\mathrm{e}_{\mathrm{m}}} \, \frac{\partial \bar{\psi}}{\bar{\mathbf{C}}_{\mathrm{e}_{\mathrm{m}}}}$, $\bar{\mathbf{\Sigma}}_{\mathrm{co}} := 2 \, \bar{\mathbf{C}}_{\mathrm{e}_{\mathrm{co}}} \, \frac{\partial \bar{\psi}}{\bar{\mathbf{C}}_{\mathrm{e}_{\mathrm{co}}}}$, $\bar{\mathbf{Y}}_{\mathrm{co}} := 2 \, \frac{\partial \bar{\psi}}{\partial \bar{\mathbf{H}}} \, \bar{\mathbf{H}}$, and $\bar{\mathbf{\Pi}}_{\mathrm{co}}  :=  2 \, \frac{\partial \bar{\psi}}{\partial \bar{\mathbf{H}}}  \cdot (\bar{\mathbf{H}} \otimes  \bar{\mathbf{H}})$.

\subsection{Evolution equations}
\label{subsec:2-4}

The next step is to introduce a set of evolution equations to describe the material behavior during the maturation process. The three phenomena considered are i) volumetric growth, ii) collagen density evolution, and iii) collagen reorientation. These phenomena and their interactions are the main driving factors influencing the evolution of the tissue's mechanical behavior. Therefore, constructing a coupled system of evolution equations is essential for a realistic description of the material response.

\subsubsection{Volumetric growth}

The term volumetric growth is often used to describe the biological process of morphogenesis. This process involves a change in the shape of the tissue, and consequently the build up of internal stresses. In this work, volumetric growth is modeled using the concept of homeostatic surface, which was first introduced by Lamm et al.\ \cite{Lamm_2022}. It was later extended by Holthusen et al.\ \cite{Holthusen_2023} to model anisotropic soft collagenous tissues. The study compared a one-surface and two-surface approach. The numerical investigations in \cite{Holthusen_2023} showed that a two-surface provides higher accuracy in modeling growth and remodeling. However, it requires introducing two material parameters for the homeostatic stresses in the matrix and collagen parts, whereas a one-surface model requires only one parameter for the total homeostatic stress.

Experimental Identification of the homeostatic stresses for each constituent can be highly challenging. Hence, no relevant investigations have been found in the literature. Therefore, we chose a one-surface approach. However, we developed a \textit{non-associative} growth model with two pseudo potentials $g_{\mathrm{m}}$ and $g_{\mathrm{co}}$ for the matrix and collagen constituents, respectively. This approach makes it is possible to define separate evolution equations for the matrix and collagen parts without the need to introduce additional material parameters for the homeostatic stress of each constituent. We examined various formulations for the homeostatic surface, and the following formulation gave us the most physiologically meaningful results:
\begin{equation}
\label{homeostatic_surface}
\phi_{\mathrm{g}} := 
\frac{1}{J^{2}} \mathrm{tr}(\mathbf{Y}_{\mathrm{g}}^{2}) + \beta_{\mathrm{g}} - (2 \, \sigma_{\mathrm{g}} \,  - \, \frac{1}{J^{2}} \mathrm{tr}(\mathbf{Y}_{\mathrm{g}}))^{2}, 
\end{equation}
where
\begin{equation}
\label{co-rotated_Kirchhoff_stress}
\tilde{\boldsymbol{\tau}} = \mathbf{U}^{-1}\, \underbrace{\left( \mathbf{U}_{\mathrm{g}_{\mathrm{m}}} \,  \bar{\mathbf{\Sigma}}_{\mathrm{m}} \, \mathbf{U}^{-1}_{\mathrm{g}_{\mathrm{m}}} + \mathbf{U}_{\mathrm{g}_{\mathrm{co}}} \,  \bar{\mathbf{\Sigma}}_{\mathrm{co}} \, \mathbf{U}^{-1}_{\mathrm{g}_{\mathrm{co}}} \right)}_{=: \mathbf{Y}_{\mathrm{g}}} \, \mathbf{U},
\end{equation}
is the co-rotated Kirchhoff stress tensor. $\tilde{\boldsymbol{\tau}}$ lives in the co-rotated configuration \textit{crc} and has the same eigenvalues as the Kirchhoff stress ${\boldsymbol{\tau}}$. Furthermore, $\mathbf{Y}_{\mathrm{g}}$ has the same eigenvalues as ${\boldsymbol{\tau}}$. Therefore, the homeostatic stress function $\phi_{\mathrm{g}}$  in Eq.\ (\ref{homeostatic_surface}) is formulated as a function of $\mathbf{Y}_{\mathrm{g}}$.

Investigation on homeostatic stress shows that it is value is influenced by the composition of biological tissue. An extensive investigation by Eichinger et al.\ \cite{Eichinger_2021} on soft collagenous tissues found a linear correlation between collagen density and homeostatic stress. It is important to consider that homeostasis is regulated locally \cite{Humphrey_2022}. These considerations are especially important when modeling the maturation process because the initial state of our system is collagen-free. In this situation considering a constant value for the homeostatic stress $\sigma_{g}$ is far from accurate. Therefore, we define the homeostatic stress as a function of the local collagen density in the ${\rho}_{\mathrm{co}}$ using the following expression:
\begin{equation}
\label{homeostatic_stress}
\sigma_{g} \, = \, \sigma_{\mathrm{g, 0}} \, \left(1 \, +  \, r_{\mathrm{1}} \, \frac{{\rho}_{\mathrm{co}}}{\rho_{\mathrm{co, f}}} \right).
\end{equation}

 Here we introduced the initial homeostatic stress $\sigma_{\mathrm{g, 0}} \, \mathrm{[MPa]}$ and the coupling coefficient $r_{\mathrm{1}} \, \mathrm{[MPa]}$ as additional material parameters. The value of $r_{\mathrm{1}}$ determines the influence of collagen density on the homeostatic stress $\sigma_{g} $.  The collagen density in the current configuration is ${\rho}_{\mathrm{co}} = \frac{1}{J} \, {\rho}^{0}_{\mathrm{co}} $ and $\rho_{\mathrm{co, f}}$ is an additional parameter representing the average collagen density in the specimen measured at the end of the maturation process.

The next step is to define the pseudo potentials for each constituent. For the matrix part, we use the following Rankine-like function
\begin{equation}
g_{\mathrm{m}} := 
\begin{cases}
\frac{1}{J} \mathrm{tr}(\mathbf{Y}_{\mathrm{g}_{\mathrm{m}}})
&  , \frac{1}{J^{2}} \, \mathrm{tr}(\mathbf{Y}_{\mathrm{g}_{\mathrm{m}}}^{2}) + \beta_{\mathrm{g}} = 0\\
\frac{1}{J} \mathrm{tr}(\mathbf{Y}_{\mathrm{g}_{\mathrm{m}}}) + \sqrt{\frac{1}{J^{2}} \mathrm{tr}(\mathbf{Y}_{\mathrm{g}_{\mathrm{m}}}) + \beta_{\mathrm{g}} }&       , else
\end{cases} ,
\end{equation}
while the growth potential in the collagen part is
\begin{equation}
g_{\mathrm{co}} := \frac{1}{J} \frac{1}{\bar{\mathbf{C}}_{\mathrm{e}_{\mathrm{co}}}:\bar{\mathbf{M}}} \, \bar{\mathbf{\Gamma}}_{\mathrm{co}} : \mathrm{sym}(\bar{\mathbf{C}}_{\mathrm{e}_{\mathrm{co}}} \bar{\mathbf{M}}).
\end{equation}

From these potential functions, we can compute the growth directions in each constituent as
\begin{equation}
\label{N_g}
\mathbf{N}_{\mathrm{g}_{\mathrm{j}}} := \frac{\partial g_{\mathrm{j}} }{\partial {\bar{\mathbf{\Sigma}}}_{\mathrm{j}} },
\end{equation}

and the normalized evolution equation is
\begin{equation}
\label{}
\bar{\mathbf{D}}_{\mathrm{g}_{\mathrm{j}}} = \dot{\gamma_{\mathrm{g}}} \,  \frac{\mathbf{N}_{\mathrm{g}_{\mathrm{j}}}}{ \lVert \mathbf{N}_{\mathrm{g}_{\mathrm{j}}} \rVert},
\end{equation}
where $\dot{\gamma_{\mathrm{g}}}$ is the growth multiplier. 

The evolution equations are based on the concept developed by Perzyna \cite{Perzyna_1971} to model visco-plastic problems. The growth multiplier is a rate quantity, which depends on the stress deviation from the homeostatic surface \cite{Lamm_2022, Holthusen_2023}. The growth multiplier  $\dot{\gamma_{\mathrm{g}}}$ is computed by solving the following equation
\begin{equation}
\label{}
\phi_{\mathrm{g}} \,  - \,  ( 4 \, \sigma_{\mathrm{g}}^{2} \,  - \, \beta_{\mathrm{g}}) (\eta_{\mathrm{g}} \, \gamma_{\mathrm{g}})^{v_{\mathrm{g}}} = 0,
\end{equation}
where $ \eta_{\mathrm{g}} $ is the growth relaxation time, and $v_{\mathrm{g}}$ is an additional parameter that describes the nonlinearity of the rate-dependent response.

\subsubsection{Collagen density evolution}

The next step is to introduce evolution equations describing changes in collagen density. Here we apply the evolution equations introduced in \cite{Sesa_2023_CBM}. The primary concept in this work is to decompose the collagen evolution into biological and mechanobiological parts as presented in the following equation
\begin{equation}
\label{eq:2-10}
\dot{\rho}^{0}_{\mathrm{co}} = \dot{\rho^{0}}_{\mathrm{bio}} + \dot{\rho}^{0}_{\mathrm{mech}}, 
\end{equation}
where the quantities $\dot{\rho}^{0}_{\mathrm{co}}$, $\dot{\rho}^{0}_{\mathrm{bio}}$ and $\dot{\rho}^{0}_{\mathrm{mech}}$ are defined in the reference configuration.

This approach was essential to make the model compatible with experimental observations which showed that unconstrained and unloaded specimens show a significant build up of collagen content during in vitro maturation. Thus, models that describe collagen densification only as a function of mechanical stimulation fail to describe this behavior. Therefore, we introduced a term to describe the biologically driven part of collagen evolution as 
\begin{equation}
\label{eq:rho_bio}
\dot{\rho}^{0}_{\mathrm{bio}} = a_{1} \, c_{\mathrm{cell}} \, \dot{\alpha}_{\mathrm{bio}},
\end{equation}
where $a_{1} \, \mathrm{[\SI{}{\micro\gram} / cells]}$ is a coefficient of the biologically-driven collagen evolution and $c_{\mathrm{cell}}$ is the cell density. The term $\dot{\alpha}_{\mathrm{bio}}$ is the time derivative of the S-shaped Weibull cumulative distribution function 
\begin{equation}
\label{eq:weibull_rate}
\dot{\alpha}_{\mathrm{bio}}  = \frac{h}{\tau}\, e^{-(t / \tau)^{h}}  \, \left(\frac{t}{\tau}\right)^{h - 1},
\end{equation}
where the parameters $\tau$ and $h$ in Eq.\ (\ref{eq:weibull_rate}) control the half-time and the steepness of the curve, respectively. For more details about the specific reasons behind choosing a Weibull cumulative distribution function, we refer the reader to \cite{Sesa_2023_CBM}.

The mechanobiologically-driven part is
\begin{equation}
\label{eq:rho_mech}
\dot{\rho}^{0}_{\mathrm{mech}}  = 
\begin{cases}
a_{2} \, c_{\mathrm{cell}} \, f_{\mathrm{mech}}  \, {\rho}^{0}_{\mathrm{co}} \, \frac{({\psi}_{\mathrm{co, m}} - {\psi}_{\mathrm{crit}})}{{\psi}_{\mathrm{crit}}} ,
&        {\psi}_{\mathrm{co, m}} \ge  {\psi}_{\mathrm{crit}},\\
0  , &       {\psi}_{\mathrm{co, m}} <  {\psi}_{\mathrm{crit}}.
\end{cases} 
\end{equation}

In Eq.\ (\ref{eq:rho_mech}), we introduced the coefficient of mechanobiological stimulation $a_{2} \, \mathrm{[mm^{3}/cells/day]}$.  Furthermore, ${\psi}_{\mathrm{co, m}}$ is the strain energy per unit mass stored in collagen fibers. The parameter ${\psi}_{\mathrm{crit}}$ is the threshold for mechanical stimulation. In addition to that, we introduced the exponential decay function $f_{\mathrm{mech}}$ which ensures collagen density increase reaches saturation level at the end of the maturation process. The exponential decay function is described by the following expression
\begin{equation}
\label{eq:exponential_decay}
f_{\mathrm{mech}} =  e^{-({\rho}^{0}_{\mathrm{co}} / \rho_{\mathrm{th}})},
\end{equation}
where the collagen saturation level is controlled by the parameter $\rho_{\mathrm{th}}$.

In our previous paper \cite{Sesa_2023_CBM}, parameters $a_{1}$, $t$, and $h$ were considered to be independent of other parameters in the case of a constrained but unloaded tissue stripe. This simplification was possible since volumetric growth was neglected. However, in this model, the evolution of the specimen shape and collagen density influence each other, since the equations are coupled.

\subsubsection{Collagen fiber reorientation}

One of the unique characteristics of living tissues is their ability to reorient their fibrous content to adapt to mechanical loading conditions. Since collagen is the main structural constituent of the tissue, the reorientation of collagen fibers significantly affect the mechanical properties of the implant. Numerical and experimental studies were done to identify the driving factors for collagen reorientation. These findings of these studies can be broadly divided into two groups, namely (i) stress-driven, and (ii) strain-driven fiber reorientation. In this work we choose a stress-driven approach. This choice is motivated by experimental observations which we will discuss in the next section.

In a stress-driven approach, collagen fibers shall reorient themselves towards the direction of the main principal Cauchy stress. It is important to consider that the vector ${\mathbf a}$  that was introduced in Eq.\ (\ref{structural_tensor}) which describes the mean orientation of a collagen bundle is defined in reference configuration, while the Cauchy stress tensor is defined in the current configuration. To overcome this challenge, we follow the approach presented in \cite{Holthusen_2023} which takes advantage of the fact that both the Cauchy stress ${\boldsymbol \sigma}$ and the co-rotated Kirchhoff stress $\tilde{\boldsymbol{\tau}}$ that we introduced in Eq.\ (\ref{co-rotated_Kirchhoff_stress}) share the same eigenvalues. This allows us to write down the evolution equation for fiber reorientation in the co-rotated configuration \textit{crc}. The relation between collagen orientation $\mathbf{a}$ in the reference configuration and $\tilde{\mathbf{a}}$ in the \textit{crc} is 
\begin{equation}
\label{n_tilde}
\tilde{\mathbf{a}} = \frac{1}{\sqrt{\mathbf{a} \cdot \mathbf{C} \cdot \mathbf{a}}} \, \mathbf{U} \,  \mathbf{n}
\end{equation}

In the next step, we identify the target orientation $\tilde{\mathbf{a}}_{\mathrm{target}}$ using the following  eigenvalue decomposition 
\begin{equation}
\label{eigen_decomposition}
\tilde{\boldsymbol{\tau}} = \sum_{i=1}^{3} \tilde{\tau}_{\mathrm{i}} \, \mathbf{a}_{\tilde{\boldsymbol{\tau}}_{\mathrm{i}}} \otimes \mathbf{a}_{\tilde{\boldsymbol{\tau}}_{\mathrm{i}}} \, = \, \sum_{i=1}^{3} {\tau}_{\mathrm{i}} \, \mathbf{a}_{\tilde{\boldsymbol{\tau}}_{\mathrm{i}}} \otimes \mathbf{a}_{\tilde{\boldsymbol{\tau}}_{\mathrm{i}}} . 
\end{equation}
since the eigenvalues of $\tilde{\boldsymbol{\tau}}$ and $\boldsymbol{\tau}$ are equal. The target orientation $\tilde{\mathbf{a}}_{target}$ is the eigenvector $\mathbf{a}_{\tilde{\boldsymbol{\tau}}_{\mathrm{i}}}$ corresponding to the maximum eigenvalue $\tau_{\mathrm{i}}$. 

The fiber reorientation towards the main principal orientation is defined using the following evolution equation
\begin{equation}
\label{evolution_reorientation}
\dot{\tilde{\mathbf{a}}} = \frac{\pi}{2 \, \eta_{\mathrm{s}}} (\tilde{\mathbf{a}}  \times \tilde{\mathbf{a}}_{\mathrm{target}}) \times \tilde{\mathbf{a}},
\end{equation}
where $ \eta_{\mathrm{s}}$ is the relaxation time for fiber reorientation \cite{Holthusen_2023}.

\subsection{Specific choices of the Helmholtz free energies}

In the previous subsections, we introduced the general form of the Helmholtz free energies. The next step is to define our specific choices of Helmholtz free energies. Finding reasonable choices for the Helmholtz free energy function to describe tissue-engineered collagenous materials was the subject of the work \cite{Sesa_2023_CBM}. In this study, we used measurements of the stress-strain behavior and the collagen fiber density at various time points during a maturation process that lasted 28 days. These data were then used to choose the Helmholtz free energy function and the corresponding material parameters. The study showed that the matrix part can be described by the following Neo-Hookean material law 
\begin{equation}
\label{Neo-Hookean}
{\psi}_{\mathrm{m}} = \frac{\mu}{2} \, (\mathrm{tr}(\bar{\mathbf{C}}_{\mathrm{e}_{\mathrm{m}}})  -3) \, - \, \mu \,  \mathrm{ln}(\bar{J}_{\mathrm{e}_{\mathrm{m}}}) + \frac{\lambda}{4}(\bar{J}^{2}_{\mathrm{e}_{\mathrm{m}}}- 1 - 2 \mathrm{ln}(\bar{J}_{\mathrm{e}_{\mathrm{m}}})),
\end{equation}
where $ \bar{J}_{\mathrm{e}_{\mathrm{m}}} = \sqrt{\mathrm{det} ( \bar{\mathbf{C}}_{\mathrm{e}_{\mathrm{m}}} )}$ represents the elastic volumetric change of the matrix part.

The collagen part is modeled using the Fung-type constitutive model \cite{Fung_1990} introduced by Holzapfel et al.\ \cite{Holzapfel_etal_2000}, leading to the following term
\begin{equation}
\label{Holzapfel-model}
{\psi}_{\mathrm{co}} = \frac{\rho^{0}_{\mathrm{co}}}{\rho_{\mathrm{co, f}}} \begin{cases} \frac{k_{1}}{2 \, k_{2}} \, (\exp{[\, \mathrm{k_{2}} \, \bar{E}_{\mathrm{co}}^2 \,]} - 1), \: \; \; \bar{E}_{\mathrm{co}} \ge 0,  \\ 0, \; \; \; \; \; \; \; \; \; \; \; \; \; \; \: \; \; \: \; \; \: \:\; \; \; \; \: \:\; \;  \; \: \:\; \; \bar{E}_{\mathrm{co}}< 0,\end{cases}
\end{equation}
where $\bar{E}_{\mathrm{co}} = \mathrm{tr}(\mathbf{\bar{\mathbf{C}}_{\mathrm{e}_{\mathrm{m}}}} \, \bar{\mathbf{H}}) - 1  $. 

The energy function in Eq.\ (\ref{Holzapfel-model}) is scaled by the relative energy density $\rho^{0}_{\mathrm{co}} / \rho_{\mathrm{co, f}}$ to take into account the influence of collagen density evolution. Such a linear correlation between collagen density $\rho^{0}_{\mathrm{co}}$ and energy ${\psi}_{\mathrm{co}} $ was identified in \cite{Sesa_2023_CBM}.

\section{Numerical implementation}
\label{sec:3}
In section \ref{sec:2} we introduced a set of evolution equations defining volumetric growth, collagen density change, and fiber reorientation. Solving this set of ordinary differential equations (ODE) requires implementing a robust time integration solution scheme. The unknown quantities that need to be solved using our solution scheme are the vectors $\mathbf{U}_{\mathrm{g}_{\mathrm{m}}}$, $\mathbf{U}_{\mathrm{g}_{\mathrm{co}}}$ and $\tilde{\mathbf{a}}$ in addition to the growth multiplier $\dot{\gamma_{\mathrm{g}}} $ and the collagen density ${\rho}^{0}_{\mathrm{co}}$. A system of ordinary differential equations is solved at each Gauss point using a fully implicit temporal integration scheme. To ensure the computational efficiency, we applied the exponential time integration algorithm developed by Vladimirov et al.\ \cite{Vladimirov_2008} to solve finite elastoplasticity problems. Such an integration scheme was later successfully applied to modeling biological growth \cite{Lamm_2022, Holthusen_2023}.

Our computational framework is implemented in the finite element program FEAP \cite{Taylor_2020}. Choosing FEAP was motivated by the possibility to easily develop user-defined material and element routines. In our implementation of the material and element routines, we relied on the automatic differentiation tool AceGen \cite{Korelc_2002, Korelc_2009}. Automatic differentiation was utilized to compute the derivative of the residual vector with respect to the internal variables, and the consistent tangent operator. 

The numerical examples presented in section \ref{sec:4} are computed using the continuum finite element formulation Q1STc \cite{Barfusz_2021a, Pacolli_2025}. Q1STc is an eight-node first-order isoparametric element. The element contains one Gauss point and applies the concept of enhanced assumed strain for hourglass stabilization. This approach eliminates volumetric and shear locking and reduces the computational cost compared to standard finite element formulations.

\section{Numerical examples}
\label{sec:4}

The next step is to evaluate our model's capabilities. The two examples explored here represent two different experimental setups. In the first example, we simulate the maturation process of a uniaxially constrained soft collagenous tissue. The setup was used by \textit{BioTex} to study the in vitro maturation process. Then, we compare our numerical with experimental data. In the second example, we study a biaxially constrained specimen under load perturbations. The experimental setup was developed by \cite{Eichinger_2020}, and later numerical computed by \cite{Holthusen_2023}, which used the experimental data to validate the in silico model. These earlier investigations primarily focused on studying stress homeostasis over 42 hours in tissues with nearly constant collagen content. This differs from our work here, where the focus is on modeling the biomechanical behavior of the tissue over a maturation process that lasts for 28 days. 

The contour and vector plots presented in this section were generated using the open-source software ParaView \cite{Ahrens_2005}. Curves and histograms were plotted using Matplotlib \cite{Hunter_2007}.

\subsection{Uniaxially constrained tissue stripe}
\label{subsec:4_1}
\begin{figure}[ht]
	\centering
	\includegraphics[scale=0.75]{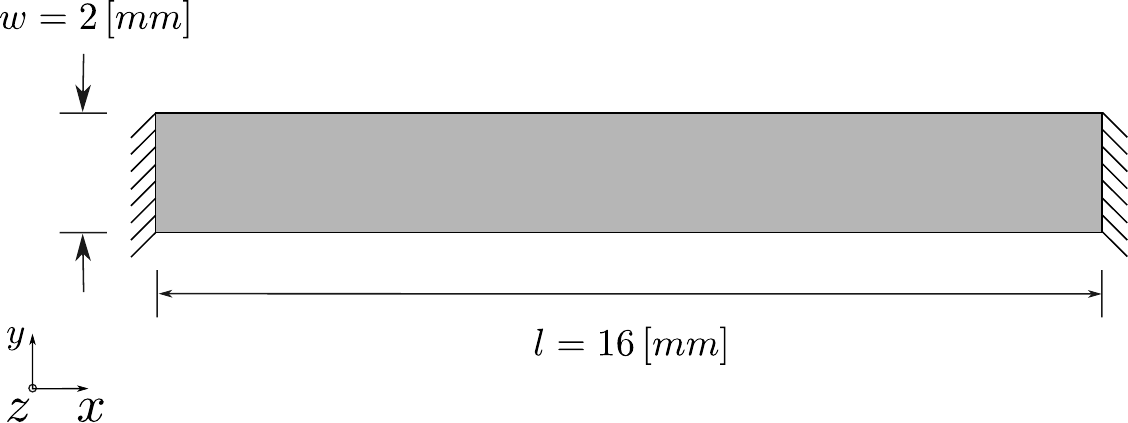}
	\caption{Schematic representation of the boundary value problem for a uniaxially constrained soft tissue construct.}
	\label{fig:uniaxial_BVP}
\end{figure}
 
In this example, we investigate a uniaxially constrained tissue stripe. The initial dimensions and boundary conditions are illustrated in Fig.\ \ref{fig:uniaxial_BVP}. The schematic shows a tissue stripe constrained from both ends. This simple geometry allows us to easily cultivate a large number of samples in a bioreactor. The mechanical and biological characteristics of the cultivated tissues are then investigated. Previous investigations \cite{Sesa_2023_CBM}, showed that scaling the energy term ${\psi}_{\mathrm{co}}$ as shown in Eq.\ (\ref{Holzapfel-model}) allows us to accurately describe the experimentally measured stress-strain response using a single set of material parameters. 
\pgfplotsset{%
	width=0.6\textwidth,
	height=0.45\textwidth
}
\begin{figure}[H]
	\centering
\begin{tikzpicture}

\definecolor{darkgray176}{RGB}{176,176,176}
\definecolor{lightgray204}{RGB}{204,204,204}

\begin{axis}[
legend cell align={left},
legend style={
  fill opacity=0.8,
  draw opacity=1,
  text opacity=1,
  at={(0.97,0.03)},
  anchor=south east,
  draw=lightgray204
},
tick align=outside,
tick pos=left,
x grid style={darkgray176},
xlabel={Time [day]},
xmin=0, xmax=30,
xtick style={color=black},
y grid style={darkgray176},
ylabel={\(\displaystyle {\rho}^{0}_{\mathrm{co}}\) [\(\displaystyle \mu {g} / mm^{3}\)]},
ymin=0, ymax=43,
ytick style={color=black}
]
\path [draw=black, semithick]
(axis cs:0,0)
--(axis cs:0,0);

\path [draw=black, semithick]
(axis cs:7,9.7836306)
--(axis cs:7,11.7799306);

\path [draw=black, semithick]
(axis cs:14,17.23186)
--(axis cs:14,31.12826);

\path [draw=black, semithick]
(axis cs:21,28.178947)
--(axis cs:21,38.868947);

\path [draw=black, semithick]
(axis cs:28,35.538)
--(axis cs:28,40.4647);

\addplot [semithick, blue]
table {%
0.1 0.022324
0.2 0.0572865
0.3 0.10266875
0.4 0.1572
0.5 0.219993125
0.6 0.29036875
0.7 0.36777125
0.8 0.451726875
0.9 0.541823125
1 0.6376875
1.1 0.738975
1.2 0.84538125
1.3 0.9566125
1.4 1.0724
1.5 1.1924625
1.6 1.316575
1.7 1.4445
1.8 1.57599375
1.9 1.71085625
2 1.848875
2.1 1.98985
2.2 2.133575
2.3 2.2799125
2.4 2.4287
2.5 2.5798125
2.6 2.7331375
2.7 2.8885875
2.8 3.04606875
2.9 3.2055375
3 3.3669125
3.1 3.53014375
3.2 3.6951625
3.3 3.861925
3.4 4.030375
3.5 4.20046875
3.6 4.3721625
3.7 4.54541875
3.8 4.7202125
3.9 4.8965125
4 5.07429375
4.1 5.2535375
4.2 5.4342125
4.3 5.6163
4.4 5.79976875
4.5 5.98458125
4.6 6.170725
4.7 6.3581875
4.8 6.546875
4.9 6.736875
5 6.9280625
5.1 7.1205
5.2 7.314125
5.3 7.5089375
5.4 7.705
5.5 7.9021875
5.6 8.1005625
5.7 8.3000625
5.8 8.50075
5.9 8.702625
6 8.905625
6.1 9.1098125
6.2 9.315125
6.3 9.521625
6.4 9.7293125
6.5 9.9381875
6.6 10.1483125
6.7 10.359625
6.8 10.5721875
6.9 10.786
7 11.0010625
7.1 11.2174375
7.2 11.4351875
7.3 11.65425
7.4 11.87475
7.5 12.096625
7.6 12.32
7.7 12.544875
7.8 12.77125
7.9 12.99925
8 13.2289375
8.1 13.46025
8.2 13.6933125
8.3 13.928125
8.4 14.1648125
8.5 14.4034375
8.6 14.6439375
8.7 14.8865
8.8 15.1311875
8.9 15.3779375
9 15.626875
9.1 15.8780625
9.2 16.1315625
9.3 16.387375
9.4 16.6455
9.5 16.906
9.6 17.1689375
9.7 17.4341875
9.8 17.7016875
9.9 17.9715
10 18.2434375
10.1 18.5173125
10.2 18.793
10.3 19.0703125
10.4 19.3488125
10.5 19.62825
10.6 19.90825
10.7 20.1883125
10.8 20.4679375
10.9 20.746625
11 21.0238125
11.1 21.299
11.2 21.5715625
11.3 21.8410625
11.4 22.107
11.5 22.3689375
11.6 22.6265
11.7 22.8794375
11.8 23.1274375
11.9 23.3703125
12 23.6079375
12.1 23.84025
12.2 24.0671875
12.3 24.2888125
12.4 24.505125
12.5 24.716125
12.6 24.922
12.7 25.1228125
12.8 25.318625
12.9 25.509625
13 25.695875
13.1 25.8775625
13.2 26.0548125
13.3 26.2278125
13.4 26.3965625
13.5 26.5613125
13.6 26.7221875
13.7 26.8793125
13.8 27.03275
13.9 27.1826875
14 27.32925
14.1 27.4725625
14.2 27.6126875
14.3 27.7498125
14.4 27.8839375
14.5 28.01525
14.6 28.143875
14.7 28.2698125
14.8 28.3931875
14.9 28.514125
15 28.6326875
15.1 28.7489375
15.2 28.8629375
15.3 28.9748125
15.4 29.084625
15.5 29.192375
15.6 29.29825
15.7 29.4021875
15.8 29.5043125
15.9 29.604625
16 29.7033125
16.1 29.80025
16.2 29.895625
16.3 29.9894375
16.4 30.08175
16.5 30.1725625
16.6 30.2619375
16.7 30.3499375
16.8 30.436625
16.9 30.522
17 30.6060625
17.1 30.6889375
17.2 30.770625
17.3 30.851125
17.4 30.9304375
17.5 31.00875
17.6 31.0859375
17.7 31.1620625
17.8 31.2371875
17.9 31.3113125
18 31.3845
18.1 31.45675
18.2 31.528125
18.3 31.5985625
18.4 31.668125
18.5 31.7368125
18.6 31.80475
18.7 31.871875
18.8 31.9381875
18.9 32.00375
19 32.0685625
19.1 32.132625
19.2 32.196
19.3 32.2586875
19.4 32.3206875
19.5 32.382
19.6 32.4426875
19.7 32.50275
19.8 32.5621875
19.9 32.6210625
20 32.6793125
20.1 32.737
20.2 32.794125
20.3 32.850625
20.4 32.9066875
20.5 32.9621875
20.6 33.017125
20.7 33.071625
20.8 33.1255625
20.9 33.1790625
21 33.2320625
21.1 33.284625
21.2 33.3366875
21.3 33.3883125
21.4 33.4395625
21.5 33.4903125
21.6 33.540625
21.7 33.5905625
21.8 33.640125
21.9 33.68925
22 33.738
22.1 33.7863125
22.2 33.8343125
22.3 33.881875
22.4 33.929125
22.5 33.976
22.6 34.0225625
22.7 34.06875
22.8 34.1145625
22.9 34.1600625
23 34.20525
23.1 34.250125
23.2 34.294625
23.3 34.338875
23.4 34.3828125
23.5 34.4264375
23.6 34.46975
23.7 34.51275
23.8 34.5555
23.9 34.5979375
24 34.640125
24.1 34.682
24.2 34.723625
24.3 34.765
24.4 34.8060625
24.5 34.8469375
24.6 34.8875
24.7 34.9278125
24.8 34.9679375
24.9 35.00775
25 35.0473125
25.1 35.0866875
25.2 35.1258125
25.3 35.1646875
25.4 35.2033125
25.5 35.24175
25.6 35.2799375
25.7 35.3179375
25.8 35.3556875
25.9 35.39325
26 35.4305625
26.1 35.4676875
26.2 35.5045625
26.3 35.5413125
26.4 35.5778125
26.5 35.614125
26.6 35.65025
26.7 35.686125
26.8 35.721875
26.9 35.757375
27 35.79275
27.1 35.827875
27.2 35.862875
27.3 35.897625
27.4 35.93225
27.5 35.9666875
27.6 36.0009375
27.7 36.035
27.8 36.0689375
27.9 36.1026875
28 36.13625
};
\addlegendentry{FEM}
\addplot [semithick, black, mark=*, mark size=2, mark options={solid}, only marks, legend image post style={sharp plot,|-|}]
table {%
0 0
7 11.0269306
14 23.45826
21 32.349947
28 38.71
};
\addlegendentry{Experimental data}
\addplot [semithick, black, mark=-, mark size=4, mark options={solid}, only marks]
table {%
0 0
7 9.7836306
14 17.23186
21 28.178947
28 35.538
};
\addplot [semithick, black, mark=-, mark size=4, mark options={solid}, only marks]
table {%
0 0
7 11.7799306
14 31.12826
21 38.868947
28 40.4647
};
\end{axis}

\end{tikzpicture}
	\caption{Collagen density evolution during the maturation process. The experimental data are measured after 7, 14, 21, and 28 days of maturation.}
	\label{fig:collagen_density_evolution}    
\end{figure}
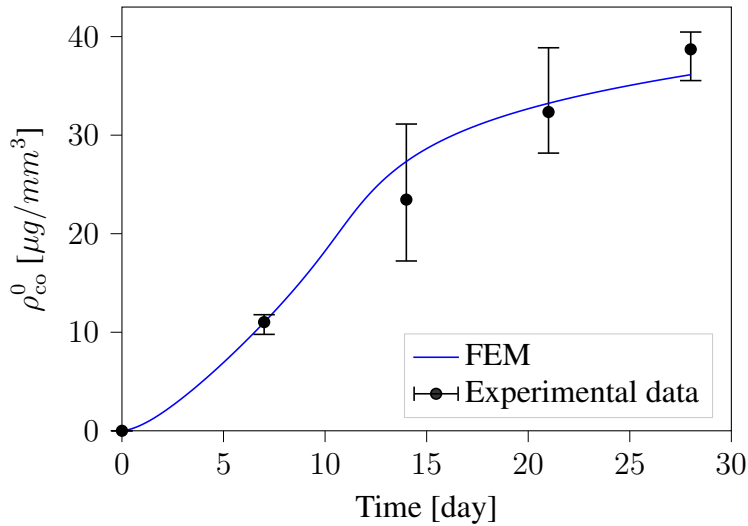

\begin{table}[H]
	\centering \renewcommand\cellalign{lc}
	\setlength{\tabcolsep}{4 pt}
	\begin{tabular}{l l l l l}
		\toprule
		Symbol & Description & Value & Units & Reference \\
		\midrule
		$ \lambda $ & First Lamé constant of matrix & $0.5$ & $\mathrm{[MPa]}$ & Selected  \\
		$ \mu  $ & shear modulus of matrix & $0.25$ & $\mathrm{[MPa]}$ & Selected  \\
		$ k_1 $ & \makecell{Stiffness-like parameter for \\ the collagen part} & $0.825$ &  $\mathrm{[MPa]}$ & \cite{Sesa_2023_CBM}  \\
		$ k_2 $ & \makecell{Exponential coefficient for  \\ the collagen part}
		& $4.0$ & $[-]$ & \cite{Sesa_2023_CBM}  \\
		$ \kappa $ & Collagen fibers dispersion parameter & $0.15$ & $[-]$ & \cite{Sesa_2023_CBM}  \\
		$ a_{1} $  & \makecell{biologically-driven collagen \\ evolution coefficient} & $1 \times 10^{-3}$ & $\mathrm{[\SI{}{\micro\gram} / cells]}$ & Fitted \\
		$ \tau $  & \makecell{Weibull cumulative distribution \\ function half-
			time} & $7$ & $\mathrm{[days]}$ & Fitted \\
		$ h  $ & \makecell{Weibull cumulative distribution \\ function parameter} & $1.65$ & $[-]$ & Fitted \\
		$ a_{2} $ &  \makecell{Mechanically-driven collagen \\ evolution coefficient} & $2.5 \times 10^{-6}$ & $\mathrm{[mm^{3}/cells/day]}$ & Fitted \\
		$ {\psi}_{\mathrm{crit}} $ & \makecell{ collagen fibers Helmholtz free energy \\ per unit  mass growth threshold} & $2 \times 10^{-5}$ & $\mathrm{[J/\SI{}{\micro\gram}]}$ & Fitted \\
		$ {\rho}_{\mathrm{th}} $ & \makecell{exponential coefficient controlling \\ collagen density saturation level} & $6.5$ & $\mathrm{[\SI{}{\micro\gram} / mm^{3}]}$ & Fitted \\
		$ {\rho}_{\mathrm{co, f}} $ & Final collagen density  & $38.7$ & $\mathrm{[\SI{}{\micro\gram} / mm^{3}]}$ & \cite{Sesa_2023_CBM} \\
		$ c_{\mathrm{cell}}  $ & Valvular interstitial cell density & $15 \times 10^{3}$ & $\mathrm{[cells/mm^{3}]}$ & \cite{Hermans_etal_2022} \\
		$\sigma_{\mathrm{g, 0}}$ & Initial homeostatic stress & $0.2 $ & $\mathrm{[MPa]}$  & Selected \\
		$r_{\mathrm{1}}$ & Homeostatic stress coupling coefficient &  $0.15 $ & $\mathrm{[MPa]}$  & Selected \\
		$\beta_{\mathrm{g}}$ & Stress-like apex parameter & $1 $ & $\mathrm{[MPa]}$  & \cite{Holthusen_2023} \\
		$\eta_{\mathrm{g}}$ & Volumetric growth relaxation time & $50 $ & $\mathrm{[days]}$  &  Fitted \\
		$\eta_{\mathrm{s}}$ & Fiber reorientation relation time  & $5 $ & $\mathrm{[days]}$  & Fitted \\
		$v_{\mathrm{g}}$ &  Perzyna exponent& $1 $ & $[-]$  & \cite{Holthusen_2023} \\
		\bottomrule
	\end{tabular}
	\caption{Material parameters for modeling the maturation of uniaxially constrained collagenous tissue.}
	\label{table:5_1}
\end{table}

Initially, the sample does not contain any collagen content. During the in vitro maturation process, we observe the synthesis of ECM which leads to increase in collagen content. Collagen density was measured using a chemical process called \textit{ hydroxyproline assay}. Experimental measurements are plotted in Fig.\ \ref{fig:collagen_density_evolution}. The black dots indicate the mean value of collagen density, and the black bars show the upper and lower range of the measurements. In \cite{Sesa_2023_CBM}, we used these experimental results to identify the parameters $\tau$ and $h$ which describe the Weibull cumulative curve. That was possible in our previous investigation because the model neglected volumetric growth and fiber reorientation, which are driven by the evolution of internal stresses. That differs significantly from the model applied here, where our system of equations takes into account the influence of volumetric growth and fiber reorientation on collagen evolution. Consequently, the computed collagen density is influenced by all model parameters listed in Table \ref{table:5_1}. In this example, we compute the average collagen density $\rho^{0}_{\mathrm{co}} $ for all elements in the computational domain. The computed results are then used to identify the parameters of our evolution equations. The plot in Fig.\ \ref{fig:collagen_density_evolution} shows that the model can accurately describe the experimental results.

\begin{figure}[H]
	\centering
	\input{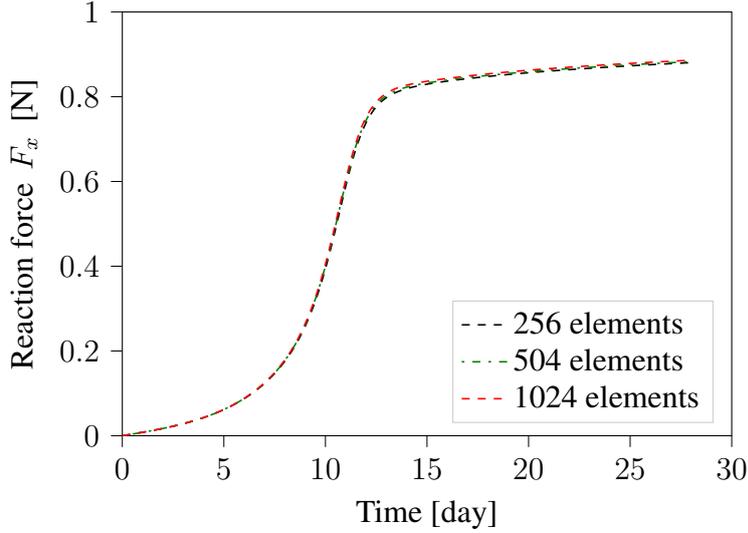}
	\caption{Evolution of the reaction force during the maturation process of uniaxially constrained tissue stripe. Computations are performed using three different mesh refinements.}
	\label{fig:mesh_convergence}    
\end{figure}

To ensure the accuracy of our finite element results, we performed a mesh convergence study. The geometry presented in Fig.\ \ref{fig:uniaxial_BVP} was discretized using three different finite element mesh refinements of 256, 504, and 1024 elements. Then we computed the boundary value problem using each mesh refinement and obtained corresponding reaction forces along the x-direction. The reaction forces computed for each mesh refinement are plotted in Fig.\ \ref{fig:mesh_convergence}. The plot shows excellent mesh convergence behavior even with a mesh of only 256 elements. The results we present in this section were computed using a mesh with 1024 elements. In this mesh, the computational domain is discretized along the x, y, and z-directions using 64, 8, and 2 elements respectively. Such a fine refinement allows us to accurately compute variations in collagen densities and orientations on the local level.

\begin{figure}[H]
	\centering
	\subfloat[ \label{fig:microscopy_images_a}]{\includegraphics[scale=0.075]{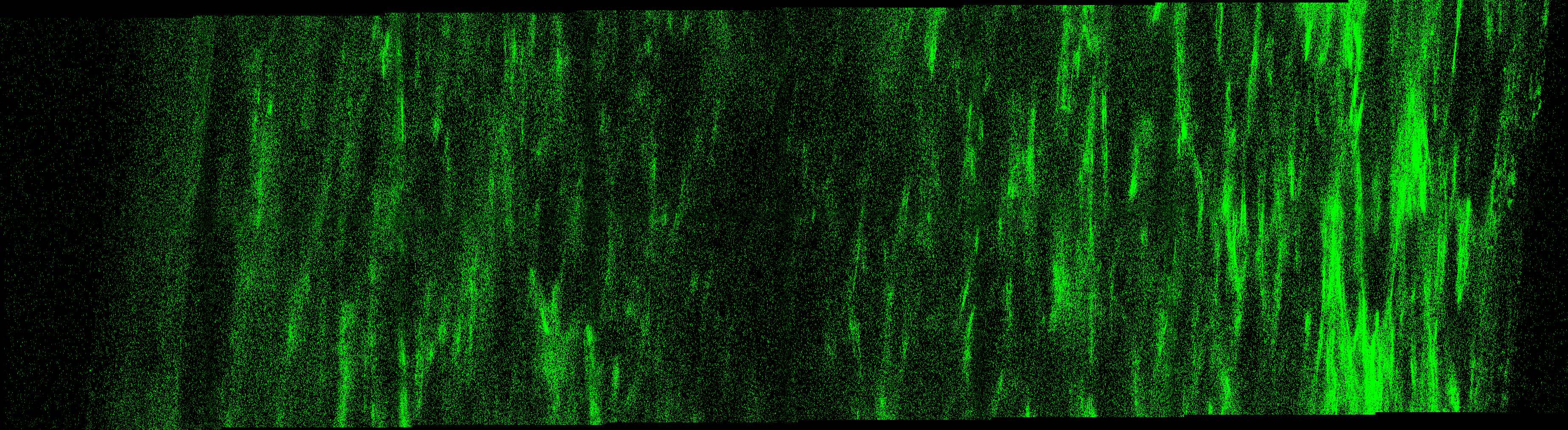}} 
    
	\vspace{0.50cm}
    
	\subfloat[\label{fig:microscopy_images_b}]{\includegraphics[scale=0.04]{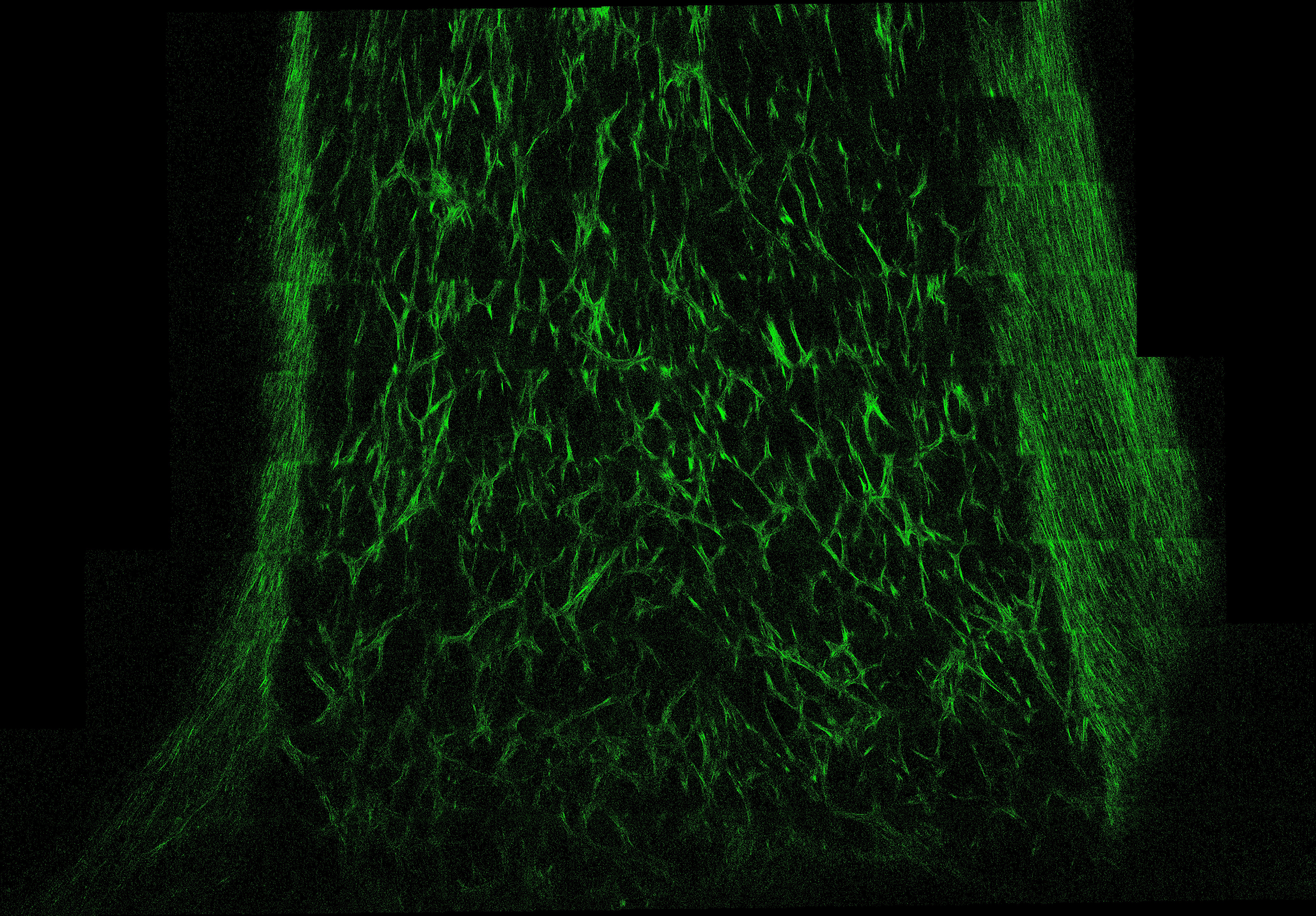}}
	
	\caption{Two-photon microscopy images for the specimen after 15 days of maturation. Collagen fibers are labeled green. (a) Collagen distribution in the middle region, and (b) in the leg region. The positions of each region in the specimen are illustrated in Fig.\ \ref{fig:fiber_orientation_areas}. }
	\label{fig:microscopy_images}    
\end{figure}

Another aspect investigated during the experiments is the evolution of the specimen shape and collagen fiber orientations. As ECM evolves, internal stresses build up within the tissue. These internal stresses alter the specimen shape, and fiber orientations. Our theoretical formulations in section \ref{sec:2} are based on the following two hypotheses (i) biological tissues seek to maintain homeostatic stress, and (ii) collagen fibers orient themselves along the main principal stress orientation. Experimental results show that the tissue width $w$ shrinks during the maturation. The behavior of collagen fibers depends on their position within the specimen. Here we study the behavior of collagen fibers along the mid-plane (middle region) and at the unixally constrained boundary (leg region). The positions of these two regions are indicated in Fig.\ \ref{fig:fiber_orientation_areas}. Furthermore, the two-photon microscopy image in Fig. \ref{fig:microscopy_images_a} shows that the collagen fibers are uniaxially oriented in the middle region, while Fig.\ \ref{fig:microscopy_images_b} shows that in the leg region, the collagen fibers orientations are highly dispersed.

\pgfplotsset{%
	width=0.6\textwidth,
	height=0.45\textwidth
}

\begin{figure}[H]
	\captionsetup[subfigure]{labelformat=empty}
	\subfloat[7 [days{]} ]{\includegraphics[scale=0.42]{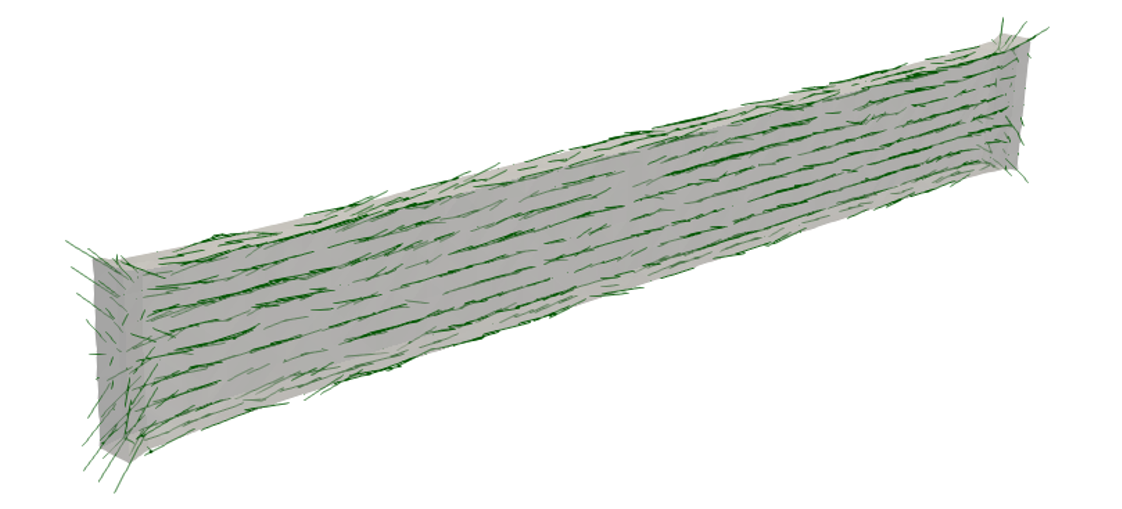}} \quad
	\subfloat[ 14 [days{]} ]{\includegraphics[scale=0.42]{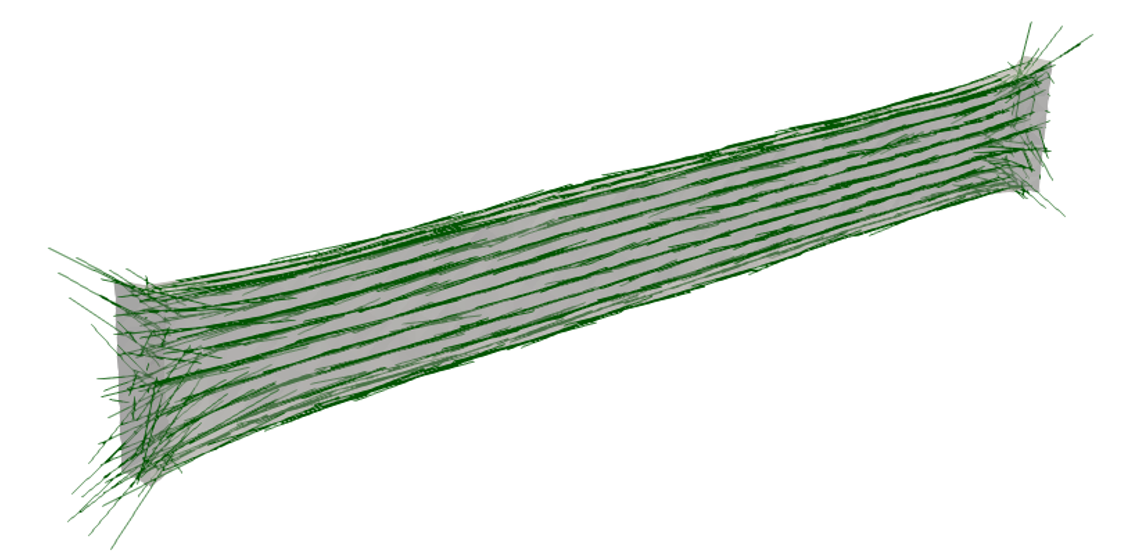}}
	
	\subfloat[ 21 [days{]} ]{\includegraphics[scale=0.42]{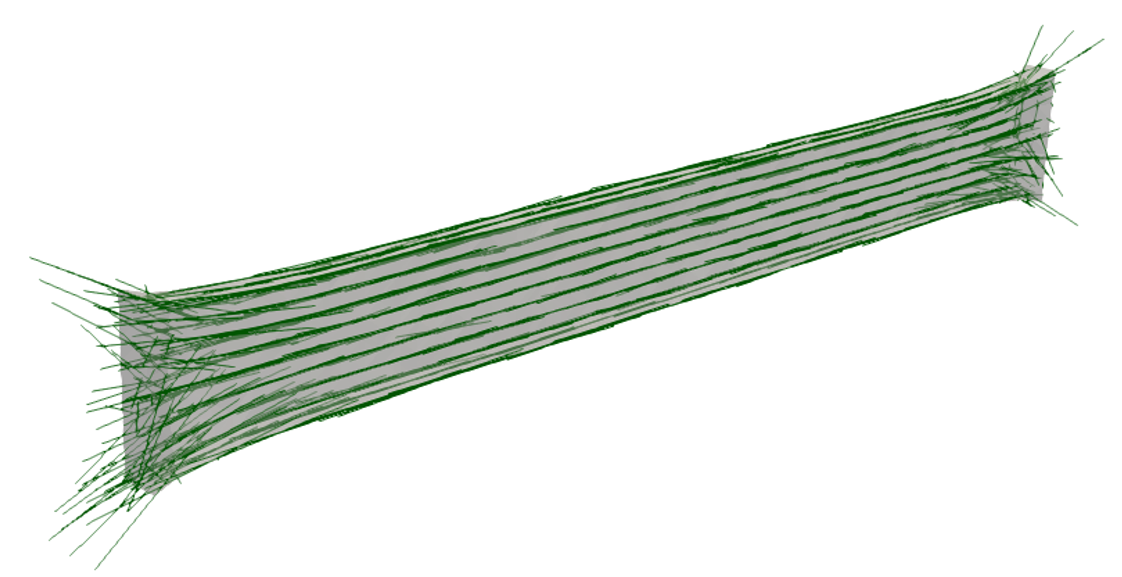}} \qquad
	\subfloat[ 28 [days{]} ]{\includegraphics[scale=0.42]{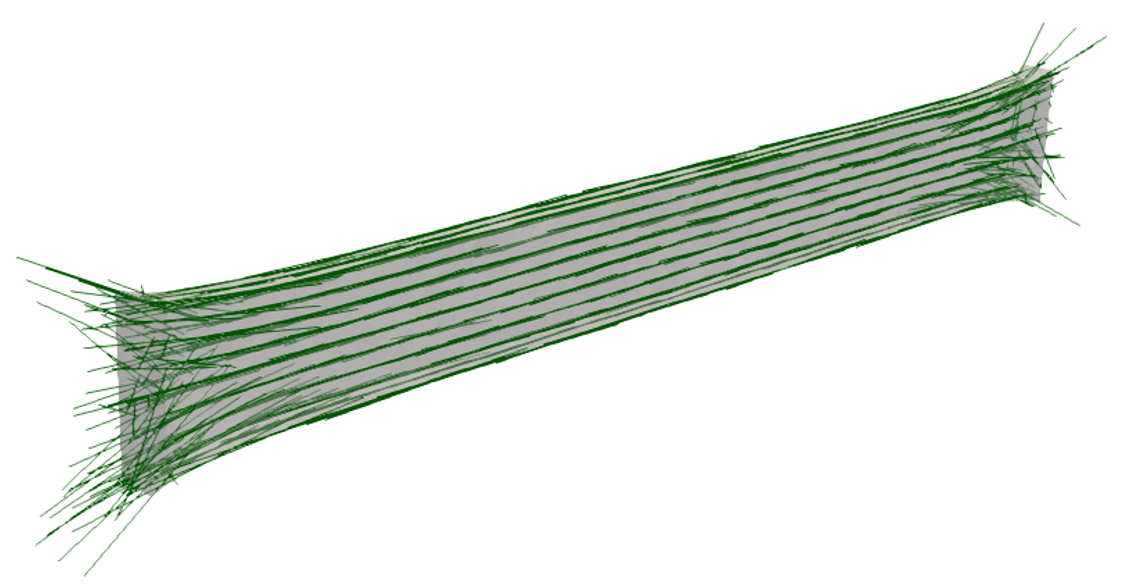}}
	
	\caption{Evolution of collagen fiber density and orientation during the maturation process for the tissue-engineered construct.}
	\label{fig:collagen_evolution}    
\end{figure}

In our computations, the initial collagen orientations are randomly defined and the initial collagen density is zero. The results in Fig.\ \ref{fig:collagen_evolution} show that the specimen shape evolves. Furthermore, collagen fibers are visualized in Fig.\ \ref{fig:collagen_evolution} using green lines, where the length of the fibers refers to the local collagen density at the corresponding element. We can observe that the density increases over time, and collagen fibers reorient themselves. Similar to microscopy images, the results obtained using our model show that collagen fibers in the middle region are uniaxially oriented, while they don't show the same level of uniaxial orientation in the leg region. These results highlight the ability of the model to describe the physical phenomena observed during the experiments.

\begin{figure}[H]
	\centering
	\captionsetup[subfigure]{labelformat=empty}
	\subfloat[ 7 [days{]} ]{\includegraphics[scale=0.45]{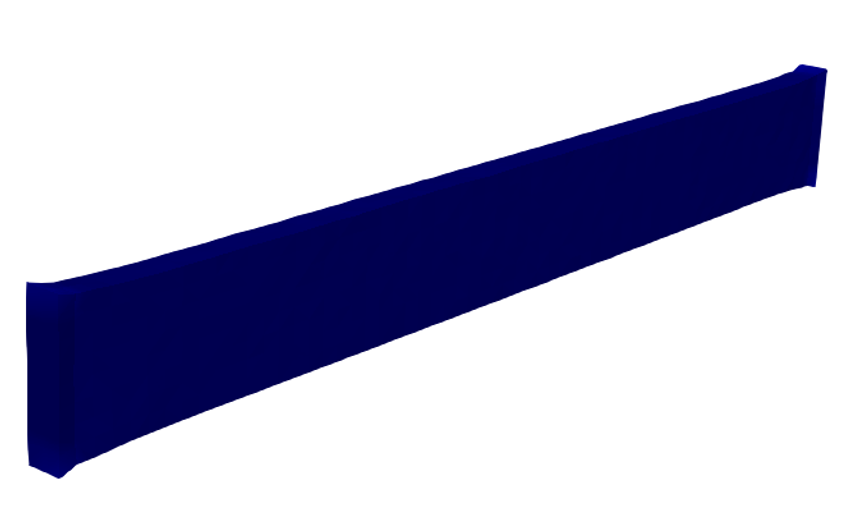}} \qquad \qquad
	\subfloat[14 [days{]}]{\includegraphics[scale=0.45]{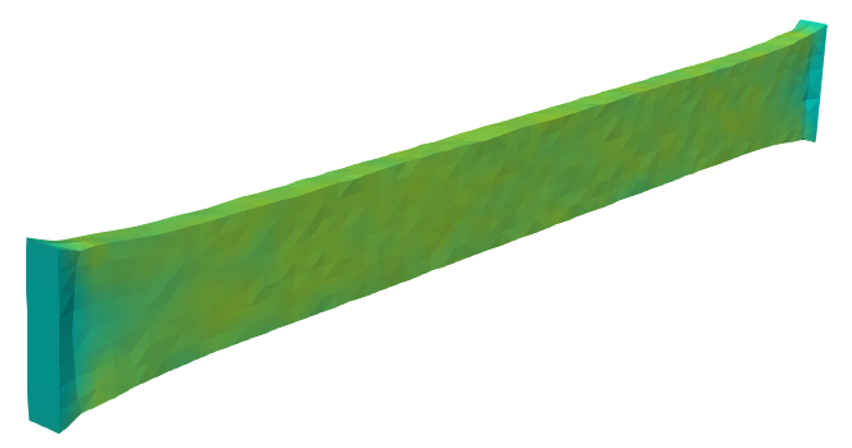}}	
	\vspace{6pt}
	
	\subfloat[21 [days{]}]{\includegraphics[scale=0.45]{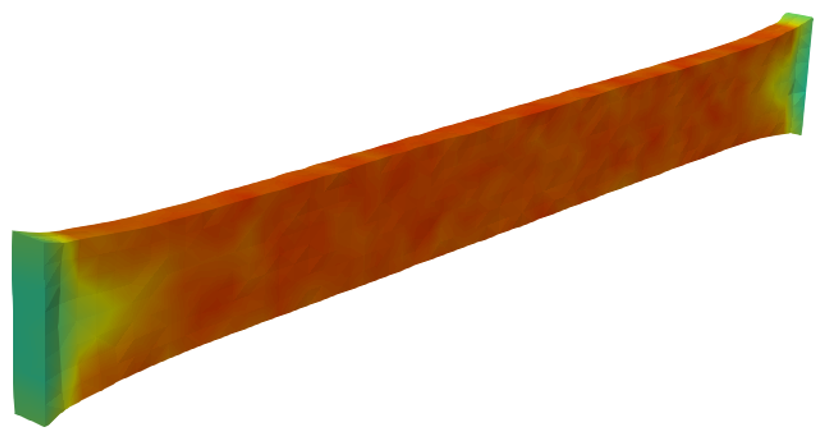}}	\qquad \qquad
	\subfloat[28 [days{]}]{\includegraphics[scale=0.45]{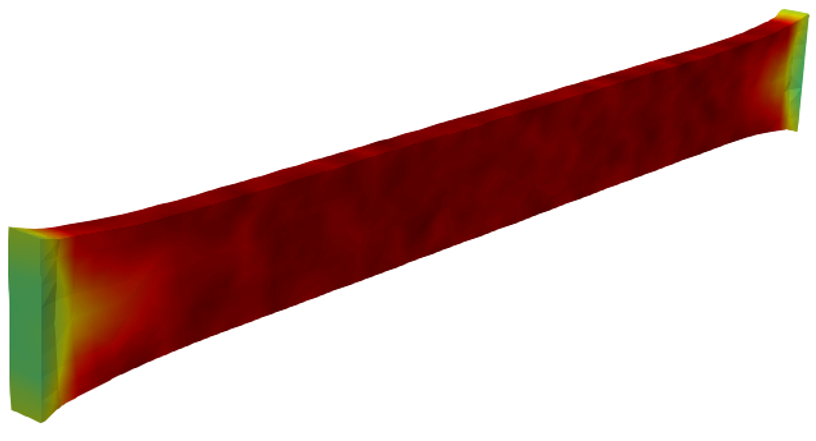}} 
	
	\subfloat{\includegraphics[scale=0.5]{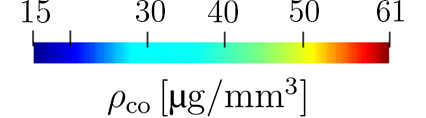}}
	
	\caption{Evolution of collagen density.}
	\label{fig:collagen_density_uniaxial}    
\end{figure}

The distribution of collagen density in the current configuration $\rho_{\mathrm{co}}$ is visualized in Fig.\ \ref{fig:collagen_density_uniaxial}. We can observe that the variation in collagen density across the sample is small. Such a result is expected since no external load is applied on the sample. Only deformation at the local level occurs due to internal stresses. Furthermore, we performed a quantitative analysis of fiber orientations at the middle and leg regions illustrated in Fig.\ \ref{fig:collagen_orientation}. Collagen fiber orientations are represented at various time steps using histograms. The two charts at the top show the initial distribution of collagen orientations which are randomly defined. The charts on the left show collagen orientations in the middle region, while the charts on the right show the orientations in the leg region. Zero degree represents a fiber orientation along the longitudinal direction. The histograms show clearly that in the middle region, collagen fibers orient themselves along the longitudinal direction, while in the leg region fiber orientation is dispersed.

The displacement boundary conditions imposed on the tissue from both sides as illustrated in Fig.\ \ref{fig:uniaxial_BVP}, lead to the build up of internal stresses within the material as it seeks to reach homeostatic stress. This behavior manifests itself in the development of reaction forces on these constrained boundaries. The results in Fig.\ \ref{fig:mesh_convergence} show a steep increase in the reaction force in the first 14 days. After that, the rate of increase in the reaction force slows down significantly. This behavior is influenced by defining homeostatic stress as a function of collagen density. This makes the model more realistic, as we know from mechanical tests that the build of ECM during the maturation process increases the stiffness of the tissue \cite{Sesa_2023_CBM}. In our numerical studies, we observed that neglecting this increase in homeostatic stress caused by the increase in collagen content leads to an initial contraction of the specimen width, followed later by an increase in width because lower levels of strain are necessary to maintain the homeostatic stress. This result contradicts experimental observations and exemplify the importance of defining homeostatic stress as a function of collagen density.

\pgfplotsset{%
	width=0.45\textwidth,
	height=0.285\textwidth
}
\begin{figure}[H]
        \centering
        \subfloat[\label{fig:fiber_orientation_areas}]{\includegraphics[scale=0.5]{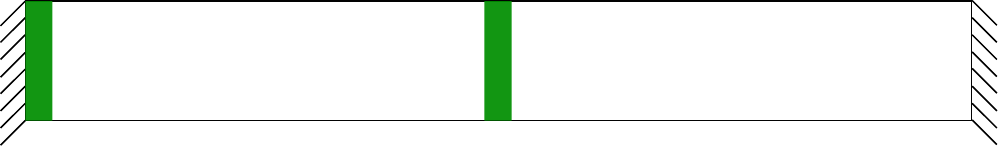}}

        \subfloat[]{\begin{tabular}[b]{c}
\begin{tikzpicture}

\definecolor{darkgray176}{RGB}{176,176,176}
\definecolor{steelblue31119180}{RGB}{31,119,180}

\begin{axis}[
tick align=outside,
tick pos=left,
x grid style={darkgray176},
xlabel={Fiber direction [degrees]},
xmin=0, xmax=90,
xtick style={color=black},
xtick={0,15,30,45,60,75,90},
xticklabels={0,15,30,45,60,75,90},
y grid style={darkgray176},
ylabel={Fiber amount [\%]},
ymin=0, ymax=3.0,
ytick style={color=black},
ytick={0,0.5,1.0,1.5,2.0,2.5,3.0},
yticklabels={0.0,0.5,1.0,1.5,2.0,2.5,3.0}
]
\draw[draw=none,fill=steelblue31119180] (axis cs:0.454292576541498,0) rectangle (axis cs:3.43873633087477,1.25966471667698);
\draw[draw=none,fill=steelblue31119180] (axis cs:3.43873633087477,0) rectangle (axis cs:6.42318008520805,1.51159766001237);
\draw[draw=none,fill=steelblue31119180] (axis cs:6.42318008520805,0) rectangle (axis cs:9.40762383954132,0.755798830006187);
\draw[draw=none,fill=steelblue31119180] (axis cs:9.40762383954132,0) rectangle (axis cs:12.3920675938746,0.503865886670791);
\draw[draw=none,fill=steelblue31119180] (axis cs:12.3920675938746,0) rectangle (axis cs:15.3765113482079,0.755798830006187);
\draw[draw=none,fill=steelblue31119180] (axis cs:15.3765113482079,0) rectangle (axis cs:18.3609551025411,2.01546354668317);
\draw[draw=none,fill=steelblue31119180] (axis cs:18.3609551025411,0) rectangle (axis cs:21.3453988568744,1.25966471667698);
\draw[draw=none,fill=steelblue31119180] (axis cs:21.3453988568744,0) rectangle (axis cs:24.3298426112077,0.503865886670791);
\draw[draw=none,fill=steelblue31119180] (axis cs:24.3298426112077,0) rectangle (axis cs:27.314286365541,2.77126237668935);
\draw[draw=none,fill=steelblue31119180] (axis cs:27.314286365541,0) rectangle (axis cs:30.2987301198743,0.755798830006186);
\draw[draw=none,fill=steelblue31119180] (axis cs:30.2987301198743,0) rectangle (axis cs:33.2831738742075,1.76353060334777);
\draw[draw=none,fill=steelblue31119180] (axis cs:33.2831738742075,0) rectangle (axis cs:36.2676176285408,0.503865886670792);
\draw[draw=none,fill=steelblue31119180] (axis cs:36.2676176285408,0) rectangle (axis cs:39.2520613828741,0.503865886670791);
\draw[draw=none,fill=steelblue31119180] (axis cs:39.2520613828741,0) rectangle (axis cs:42.2365051372074,1.25966471667698);
\draw[draw=none,fill=steelblue31119180] (axis cs:42.2365051372074,0) rectangle (axis cs:45.2209488915406,0.755798830006188);
\draw[draw=none,fill=steelblue31119180] (axis cs:45.2209488915406,0) rectangle (axis cs:48.2053926458739,0.755798830006186);
\draw[draw=none,fill=steelblue31119180] (axis cs:48.2053926458739,0) rectangle (axis cs:51.1898364002072,0.755798830006186);
\draw[draw=none,fill=steelblue31119180] (axis cs:51.1898364002072,0) rectangle (axis cs:54.1742801545404,1.51159766001238);
\draw[draw=none,fill=steelblue31119180] (axis cs:54.1742801545404,0) rectangle (axis cs:57.1587239088737,1.00773177334158);
\draw[draw=none,fill=steelblue31119180] (axis cs:57.1587239088737,0) rectangle (axis cs:60.143167663207,1.00773177334158);
\draw[draw=none,fill=steelblue31119180] (axis cs:60.143167663207,0) rectangle (axis cs:63.1276114175403,1.25966471667698);
\draw[draw=none,fill=steelblue31119180] (axis cs:63.1276114175403,0) rectangle (axis cs:66.1120551718736,0.755798830006183);
\draw[draw=none,fill=steelblue31119180] (axis cs:66.1120551718736,0) rectangle (axis cs:69.0964989262068,1.00773177334158);
\draw[draw=none,fill=steelblue31119180] (axis cs:69.0964989262068,0) rectangle (axis cs:72.0809426805401,1.76353060334777);
\draw[draw=none,fill=steelblue31119180] (axis cs:72.0809426805401,0) rectangle (axis cs:75.0653864348734,0.50386588667079);
\draw[draw=none,fill=steelblue31119180] (axis cs:75.0653864348734,0) rectangle (axis cs:78.0498301892067,0.755798830006188);
\draw[draw=none,fill=steelblue31119180] (axis cs:78.0498301892067,0) rectangle (axis cs:81.0342739435399,1.51159766001238);
\draw[draw=none,fill=steelblue31119180] (axis cs:81.0342739435399,0) rectangle (axis cs:84.0187176978732,0.755798830006184);
\draw[draw=none,fill=steelblue31119180] (axis cs:84.0187176978732,0) rectangle (axis cs:87.0031614522065,1.00773177334158);
\draw[draw=none,fill=steelblue31119180] (axis cs:87.0031614522065,0) rectangle (axis cs:89.9876052065398,2.26739649001856);
\draw (axis cs:46,2.55) node[
scale=1.25,
text=black,
rotate=0.0
]{ $t = 0$ days};
\end{axis}

\end{tikzpicture}
        \
\begin{tikzpicture}

\definecolor{darkgray176}{RGB}{176,176,176}
\definecolor{steelblue31119180}{RGB}{31,119,180}

\begin{axis}[
tick align=outside,
tick pos=left,
x grid style={darkgray176},
xlabel={Fiber direction [degrees]},
xmin=0, xmax=90,
xtick style={color=black},
xtick={0,15,30,45,60,75,90},
xticklabels={0,15,30,45,60,75,90},
y grid style={darkgray176},
ylabel={Fiber amount [\%]},
ymin=0, ymax=3.0,
ytick style={color=black},
ytick={0,0.5,1.0,1.5,2.0,2.5,3.0},
yticklabels={0.0,0.5,1.0,1.5,2.0,2.5,3.0}
]
\draw[draw=none,fill=steelblue31119180] (axis cs:1.09910152806917,0) rectangle (axis cs:4.02861618554162,2.7392562578773);
\draw[draw=none,fill=steelblue31119180] (axis cs:4.02861618554162,0) rectangle (axis cs:6.95813084301407,1.05356009918358);
\draw[draw=none,fill=steelblue31119180] (axis cs:6.95813084301407,0) rectangle (axis cs:9.88764550048652,1.47498413885701);
\draw[draw=none,fill=steelblue31119180] (axis cs:9.88764550048652,0) rectangle (axis cs:12.817160157959,0.421424039673431);
\draw[draw=none,fill=steelblue31119180] (axis cs:12.817160157959,0) rectangle (axis cs:15.7466748154314,1.2642721190203);
\draw[draw=none,fill=steelblue31119180] (axis cs:15.7466748154314,0) rectangle (axis cs:18.6761894729039,1.05356009918358);
\draw[draw=none,fill=steelblue31119180] (axis cs:18.6761894729039,0) rectangle (axis cs:21.6057041303763,0.842848079346864);
\draw[draw=none,fill=steelblue31119180] (axis cs:21.6057041303763,0) rectangle (axis cs:24.5352187878488,1.89640817853044);
\draw[draw=none,fill=steelblue31119180] (axis cs:24.5352187878488,0) rectangle (axis cs:27.4647334453212,1.47498413885701);
\draw[draw=none,fill=steelblue31119180] (axis cs:27.4647334453212,0) rectangle (axis cs:30.3942481027937,0.421424039673432);
\draw[draw=none,fill=steelblue31119180] (axis cs:30.3942481027937,0) rectangle (axis cs:33.3237627602661,1.47498413885701);
\draw[draw=none,fill=steelblue31119180] (axis cs:33.3237627602661,0) rectangle (axis cs:36.2532774177386,0.842848079346862);
\draw[draw=none,fill=steelblue31119180] (axis cs:36.2532774177386,0) rectangle (axis cs:39.182792075211,1.2642721190203);
\draw[draw=none,fill=steelblue31119180] (axis cs:39.182792075211,0) rectangle (axis cs:42.1123067326835,0.421424039673432);
\draw[draw=none,fill=steelblue31119180] (axis cs:42.1123067326834,0) rectangle (axis cs:45.0418213901559,1.47498413885701);
\draw[draw=none,fill=steelblue31119180] (axis cs:45.0418213901559,0) rectangle (axis cs:47.9713360476283,1.47498413885701);
\draw[draw=none,fill=steelblue31119180] (axis cs:47.9713360476283,0) rectangle (axis cs:50.9008507051008,0.210712019836716);
\draw[draw=none,fill=steelblue31119180] (axis cs:50.9008507051008,0) rectangle (axis cs:53.8303653625732,1.26427211902029);
\draw[draw=none,fill=steelblue31119180] (axis cs:53.8303653625732,0) rectangle (axis cs:56.7598800200457,1.05356009918358);
\draw[draw=none,fill=steelblue31119180] (axis cs:56.7598800200457,0) rectangle (axis cs:59.6893946775181,0.842848079346864);
\draw[draw=none,fill=steelblue31119180] (axis cs:59.6893946775181,0) rectangle (axis cs:62.6189093349906,0.421424039673431);
\draw[draw=none,fill=steelblue31119180] (axis cs:62.6189093349906,0) rectangle (axis cs:65.548423992463,1.47498413885702);
\draw[draw=none,fill=steelblue31119180] (axis cs:65.548423992463,0) rectangle (axis cs:68.4779386499355,0.632136059510146);
\draw[draw=none,fill=steelblue31119180] (axis cs:68.4779386499355,0) rectangle (axis cs:71.4074533074079,1.26427211902029);
\draw[draw=none,fill=steelblue31119180] (axis cs:71.4074533074079,0) rectangle (axis cs:74.3369679648804,0.632136059510149);
\draw[draw=none,fill=steelblue31119180] (axis cs:74.3369679648804,0) rectangle (axis cs:77.2664826223528,1.89640817853044);
\draw[draw=none,fill=steelblue31119180] (axis cs:77.2664826223528,0) rectangle (axis cs:80.1959972798253,1.05356009918358);
\draw[draw=none,fill=steelblue31119180] (axis cs:80.1959972798253,0) rectangle (axis cs:83.1255119372977,1.89640817853045);
\draw[draw=none,fill=steelblue31119180] (axis cs:83.1255119372977,0) rectangle (axis cs:86.0550265947702,0.842848079346862);
\draw[draw=none,fill=steelblue31119180] (axis cs:86.0550265947702,0) rectangle (axis cs:88.9845412522426,1.05356009918358);
\draw (axis cs:45,2.55) node[
scale=1.25,
text=black,
rotate=0.0
]{ $t = 0$ days};
\end{axis}

\end{tikzpicture}\\   

\begin{tikzpicture}

\definecolor{darkgray176}{RGB}{176,176,176}
\definecolor{steelblue31119180}{RGB}{31,119,180}

\begin{axis}[
tick align=outside,
tick pos=left,
x grid style={darkgray176},
xlabel={Fiber direction [degrees]},
xmin=0, xmax=90,
xtick style={color=black},
xtick={0,15,30,45,60,75,90},
xticklabels={0,15,30,45,60,75,90},
y grid style={darkgray176},
ylabel={Fiber amount [\%]},
ymin=0, ymax=10.0,
ytick style={color=black},
ytick={0,2,4,6,8,10},
yticklabels={0.0,2.0,4.0,6.0,8.0,10.0}
]
\draw[draw=none,fill=steelblue31119180] (axis cs:0.0150998145906496,0) rectangle (axis cs:3.00887524413386,8.95578002889464);
\draw[draw=none,fill=steelblue31119180] (axis cs:3.00887524413386,0) rectangle (axis cs:6.00265067367707,6.29325083111516);
\draw[draw=none,fill=steelblue31119180] (axis cs:6.00265067367707,0) rectangle (axis cs:8.99642610322027,3.38867352444662);
\draw[draw=none,fill=steelblue31119180] (axis cs:8.99642610322027,0) rectangle (axis cs:11.9902015327635,1.93638487111236);
\draw[draw=none,fill=steelblue31119180] (axis cs:11.9902015327635,0) rectangle (axis cs:14.9839769623067,0.726144326667133);
\draw[draw=none,fill=steelblue31119180] (axis cs:14.9839769623067,0) rectangle (axis cs:17.9777523918499,0.48409621777809);
\draw[draw=none,fill=steelblue31119180] (axis cs:17.9777523918499,0) rectangle (axis cs:20.9715278213931,0.242048108889044);
\draw[draw=none,fill=steelblue31119180] (axis cs:20.9715278213931,0) rectangle (axis cs:23.9653032509363,0.484096217778089);
\draw[draw=none,fill=steelblue31119180] (axis cs:23.9653032509363,0) rectangle (axis cs:26.9590786804795,0);
\draw[draw=none,fill=steelblue31119180] (axis cs:26.9590786804795,0) rectangle (axis cs:29.9528541100227,0.242048108889044);
\draw[draw=none,fill=steelblue31119180] (axis cs:29.9528541100227,0) rectangle (axis cs:32.9466295395659,0);
\draw[draw=none,fill=steelblue31119180] (axis cs:32.9466295395659,0) rectangle (axis cs:35.9404049691091,0);
\draw[draw=none,fill=steelblue31119180] (axis cs:35.9404049691092,0) rectangle (axis cs:38.9341803986524,0);
\draw[draw=none,fill=steelblue31119180] (axis cs:38.9341803986524,0) rectangle (axis cs:41.9279558281956,0);
\draw[draw=none,fill=steelblue31119180] (axis cs:41.9279558281956,0) rectangle (axis cs:44.9217312577388,0);
\draw[draw=none,fill=steelblue31119180] (axis cs:44.9217312577388,0) rectangle (axis cs:47.915506687282,0);
\draw[draw=none,fill=steelblue31119180] (axis cs:47.915506687282,0) rectangle (axis cs:50.9092821168252,0);
\draw[draw=none,fill=steelblue31119180] (axis cs:50.9092821168252,0) rectangle (axis cs:53.9030575463684,0);
\draw[draw=none,fill=steelblue31119180] (axis cs:53.9030575463684,0) rectangle (axis cs:56.8968329759116,0);
\draw[draw=none,fill=steelblue31119180] (axis cs:56.8968329759116,0) rectangle (axis cs:59.8906084054548,0);
\draw[draw=none,fill=steelblue31119180] (axis cs:59.8906084054548,0) rectangle (axis cs:62.884383834998,0);
\draw[draw=none,fill=steelblue31119180] (axis cs:62.884383834998,0) rectangle (axis cs:65.8781592645412,0);
\draw[draw=none,fill=steelblue31119180] (axis cs:65.8781592645412,0) rectangle (axis cs:68.8719346940844,0.242048108889045);
\draw[draw=none,fill=steelblue31119180] (axis cs:68.8719346940844,0) rectangle (axis cs:71.8657101236277,0);
\draw[draw=none,fill=steelblue31119180] (axis cs:71.8657101236277,0) rectangle (axis cs:74.8594855531709,0);
\draw[draw=none,fill=steelblue31119180] (axis cs:74.8594855531709,0) rectangle (axis cs:77.8532609827141,0.484096217778089);
\draw[draw=none,fill=steelblue31119180] (axis cs:77.8532609827141,0) rectangle (axis cs:80.8470364122573,0.242048108889045);
\draw[draw=none,fill=steelblue31119180] (axis cs:80.8470364122573,0) rectangle (axis cs:83.8408118418005,1.45228865333427);
\draw[draw=none,fill=steelblue31119180] (axis cs:83.8408118418005,0) rectangle (axis cs:86.8345872713437,2.90457730666854);
\draw[draw=none,fill=steelblue31119180] (axis cs:86.8345872713437,0) rectangle (axis cs:89.8283627008869,5.32505839555896);
\draw (axis cs:45,8.5) node[
scale=1.25,
text=black,
rotate=0.0
]{ $t = 7$ days};
\end{axis}
\end{tikzpicture}
  	\
\begin{tikzpicture}

\definecolor{darkgray176}{RGB}{176,176,176}
\definecolor{steelblue31119180}{RGB}{31,119,180}

\begin{axis}[
tick align=outside,
tick pos=left,
x grid style={darkgray176},
xlabel={Fiber direction [degrees]},
xmin=0, xmax=90,
xtick style={color=black},
xtick={0,15,30,45,60,75,90},
xticklabels={0,15,30,45,60,75,90},
y grid style={darkgray176},
ylabel={Fiber amount [\%]},
ymin=0, ymax=7.0,
ytick style={color=black},
ytick={0,1,2,3,4,5,6,7},
yticklabels={0.0,1.0,2.0,3.0,4.0,5.0,6.0,7.0}
]
\draw[draw=none,fill=steelblue31119180] (axis cs:0.0212134184288424,0) rectangle (axis cs:3.00757706282338,6.52315719054671);
\draw[draw=none,fill=steelblue31119180] (axis cs:3.00757706282338,0) rectangle (axis cs:5.99394070721791,3.26157859527335);
\draw[draw=none,fill=steelblue31119180] (axis cs:5.99394070721791,0) rectangle (axis cs:8.98030435161245,1.30463143810934);
\draw[draw=none,fill=steelblue31119180] (axis cs:8.98030435161245,0) rectangle (axis cs:11.966667996007,1.52207001112757);
\draw[draw=none,fill=steelblue31119180] (axis cs:11.966667996007,0) rectangle (axis cs:14.9530316404015,0.869754292072894);
\draw[draw=none,fill=steelblue31119180] (axis cs:14.9530316404015,0) rectangle (axis cs:17.939395284796,1.08719286509112);
\draw[draw=none,fill=steelblue31119180] (axis cs:17.939395284796,0) rectangle (axis cs:20.9257589291906,1.30463143810934);
\draw[draw=none,fill=steelblue31119180] (axis cs:20.9257589291906,0) rectangle (axis cs:23.9121225735851,0.434877146036447);
\draw[draw=none,fill=steelblue31119180] (axis cs:23.9121225735851,0) rectangle (axis cs:26.8984862179797,0.869754292072895);
\draw[draw=none,fill=steelblue31119180] (axis cs:26.8984862179797,0) rectangle (axis cs:29.8848498623742,0.65231571905467);
\draw[draw=none,fill=steelblue31119180] (axis cs:29.8848498623742,0) rectangle (axis cs:32.8712135067687,0.434877146036447);
\draw[draw=none,fill=steelblue31119180] (axis cs:32.8712135067687,0) rectangle (axis cs:35.8575771511633,1.30463143810934);
\draw[draw=none,fill=steelblue31119180] (axis cs:35.8575771511633,0) rectangle (axis cs:38.8439407955578,0);
\draw[draw=none,fill=steelblue31119180] (axis cs:38.8439407955578,0) rectangle (axis cs:41.8303044399523,0.869754292072894);
\draw[draw=none,fill=steelblue31119180] (axis cs:41.8303044399523,0) rectangle (axis cs:44.8166680843469,0.217438573018224);
\draw[draw=none,fill=steelblue31119180] (axis cs:44.8166680843469,0) rectangle (axis cs:47.8030317287414,0.65231571905467);
\draw[draw=none,fill=steelblue31119180] (axis cs:47.8030317287414,0) rectangle (axis cs:50.7893953731359,0.217438573018223);
\draw[draw=none,fill=steelblue31119180] (axis cs:50.7893953731359,0) rectangle (axis cs:53.7757590175305,0.652315719054672);
\draw[draw=none,fill=steelblue31119180] (axis cs:53.7757590175305,0) rectangle (axis cs:56.762122661925,0.434877146036447);
\draw[draw=none,fill=steelblue31119180] (axis cs:56.762122661925,0) rectangle (axis cs:59.7484863063195,0.65231571905467);
\draw[draw=none,fill=steelblue31119180] (axis cs:59.7484863063195,0) rectangle (axis cs:62.7348499507141,0);
\draw[draw=none,fill=steelblue31119180] (axis cs:62.7348499507141,0) rectangle (axis cs:65.7212135951086,0.217438573018223);
\draw[draw=none,fill=steelblue31119180] (axis cs:65.7212135951086,0) rectangle (axis cs:68.7075772395031,0.217438573018224);
\draw[draw=none,fill=steelblue31119180] (axis cs:68.7075772395031,0) rectangle (axis cs:71.6939408838977,0.652315719054672);
\draw[draw=none,fill=steelblue31119180] (axis cs:71.6939408838977,0) rectangle (axis cs:74.6803045282922,0.869754292072892);
\draw[draw=none,fill=steelblue31119180] (axis cs:74.6803045282922,0) rectangle (axis cs:77.6666681726867,1.08719286509112);
\draw[draw=none,fill=steelblue31119180] (axis cs:77.6666681726867,0) rectangle (axis cs:80.6530318170813,1.08719286509112);
\draw[draw=none,fill=steelblue31119180] (axis cs:80.6530318170813,0) rectangle (axis cs:83.6393954614758,1.73950858414578);
\draw[draw=none,fill=steelblue31119180] (axis cs:83.6393954614758,0) rectangle (axis cs:86.6257591058703,2.17438573018224);
\draw[draw=none,fill=steelblue31119180] (axis cs:86.6257591058703,0) rectangle (axis cs:89.6121227502649,2.17438573018224);
\draw (axis cs:45,5.95) node[
scale=1.25,
text=black,
rotate=0.0
]{ $t = 7$ days};
\end{axis}

\end{tikzpicture}\\
	
\begin{tikzpicture}

\definecolor{darkgray176}{RGB}{176,176,176}
\definecolor{steelblue31119180}{RGB}{31,119,180}

\begin{axis}[
tick align=outside,
tick pos=left,
x grid style={darkgray176},
xlabel={Fiber direction [degrees]},
xmin=0, xmax=90,
xtick style={color=black},
xtick={0,15,30,45,60,75,90},
xticklabels={0,15,30,45,60,75,90},
y grid style={darkgray176},
ylabel={Fiber amount [\%]},
ymin=0, ymax=20.0,
ytick style={color=black},
ytick={0.0,4.0,8.0,12.0,16.0,20.0},
yticklabels={0.0,4.0,8.0,12.0,16.0,20.0}
]
\draw[draw=none,fill=steelblue31119180] (axis cs:0.0030367486235976,0) rectangle (axis cs:3.00037221429511,19.6120135557179);
\draw[draw=none,fill=steelblue31119180] (axis cs:3.00037221429511,0) rectangle (axis cs:5.99770767996663,5.18478519289094);
\draw[draw=none,fill=steelblue31119180] (axis cs:5.99770767996663,0) rectangle (axis cs:8.99504314563815,0.676276329507513);
\draw[draw=none,fill=steelblue31119180] (axis cs:8.99504314563815,0) rectangle (axis cs:11.9923786113097,1.35255265901503);
\draw[draw=none,fill=steelblue31119180] (axis cs:11.9923786113097,0) rectangle (axis cs:14.9897140769812,0.225425443169171);
\draw[draw=none,fill=steelblue31119180] (axis cs:14.9897140769812,0) rectangle (axis cs:17.9870495426527,0);
\draw[draw=none,fill=steelblue31119180] (axis cs:17.9870495426527,0) rectangle (axis cs:20.9843850083242,0);
\draw[draw=none,fill=steelblue31119180] (axis cs:20.9843850083242,0) rectangle (axis cs:23.9817204739957,0);
\draw[draw=none,fill=steelblue31119180] (axis cs:23.9817204739957,0) rectangle (axis cs:26.9790559396673,0.450850886338342);
\draw[draw=none,fill=steelblue31119180] (axis cs:26.9790559396673,0) rectangle (axis cs:29.9763914053388,0);
\draw[draw=none,fill=steelblue31119180] (axis cs:29.9763914053388,0) rectangle (axis cs:32.9737268710103,0);
\draw[draw=none,fill=steelblue31119180] (axis cs:32.9737268710103,0) rectangle (axis cs:35.9710623366818,0);
\draw[draw=none,fill=steelblue31119180] (axis cs:35.9710623366818,0) rectangle (axis cs:38.9683978023533,0);
\draw[draw=none,fill=steelblue31119180] (axis cs:38.9683978023533,0) rectangle (axis cs:41.9657332680248,0.225425443169172);
\draw[draw=none,fill=steelblue31119180] (axis cs:41.9657332680248,0) rectangle (axis cs:44.9630687336964,0);
\draw[draw=none,fill=steelblue31119180] (axis cs:44.9630687336964,0) rectangle (axis cs:47.9604041993679,0.225425443169171);
\draw[draw=none,fill=steelblue31119180] (axis cs:47.9604041993679,0) rectangle (axis cs:50.9577396650394,0);
\draw[draw=none,fill=steelblue31119180] (axis cs:50.9577396650394,0) rectangle (axis cs:53.9550751307109,0);
\draw[draw=none,fill=steelblue31119180] (axis cs:53.9550751307109,0) rectangle (axis cs:56.9524105963824,0);
\draw[draw=none,fill=steelblue31119180] (axis cs:56.9524105963824,0) rectangle (axis cs:59.9497460620539,0);
\draw[draw=none,fill=steelblue31119180] (axis cs:59.9497460620539,0) rectangle (axis cs:62.9470815277255,0);
\draw[draw=none,fill=steelblue31119180] (axis cs:62.9470815277255,0) rectangle (axis cs:65.944416993397,0);
\draw[draw=none,fill=steelblue31119180] (axis cs:65.944416993397,0) rectangle (axis cs:68.9417524590685,0);
\draw[draw=none,fill=steelblue31119180] (axis cs:68.9417524590685,0) rectangle (axis cs:71.93908792474,0);
\draw[draw=none,fill=steelblue31119180] (axis cs:71.93908792474,0) rectangle (axis cs:74.9364233904115,0);
\draw[draw=none,fill=steelblue31119180] (axis cs:74.9364233904115,0) rectangle (axis cs:77.933758856083,0);
\draw[draw=none,fill=steelblue31119180] (axis cs:77.933758856083,0) rectangle (axis cs:80.9310943217546,0);
\draw[draw=none,fill=steelblue31119180] (axis cs:80.9310943217546,0) rectangle (axis cs:83.9284297874261,0);
\draw[draw=none,fill=steelblue31119180] (axis cs:83.9284297874261,0) rectangle (axis cs:86.9257652530976,1.12712721584585);
\draw[draw=none,fill=steelblue31119180] (axis cs:86.9257652530976,0) rectangle (axis cs:89.9231007187691,4.28308342021426);
\draw (axis cs:45,17) node[
scale=1.25,
text=black,
rotate=0.0
]{ $t = 14$ days};
\end{axis}

\end{tikzpicture}
	  \
\begin{tikzpicture}

\definecolor{darkgray176}{RGB}{176,176,176}
\definecolor{steelblue31119180}{RGB}{31,119,180}

\begin{axis}[
tick align=outside,
tick pos=left,
x grid style={darkgray176},
xlabel={Fiber direction [degrees]},
xmin=0, xmax=90,
xtick style={color=black},
xtick={0,15,30,45,60,75,90},
xticklabels={0,15,30,45,60,75,90},
y grid style={darkgray176},
ylabel={Fiber amount [\%]},
ymin=0, ymax=4.0,
ytick style={color=black},
ytick={0,1.0,2.0,3.0,4.0},
yticklabels={0,1.0,2.0,3.0,4.0}
]
\draw[draw=none,fill=steelblue31119180] (axis cs:0.0276483221611097,0) rectangle (axis cs:3.00934856019751,3.94563669824187);
\draw[draw=none,fill=steelblue31119180] (axis cs:3.00934856019751,0) rectangle (axis cs:5.99104879823391,2.2546495418525);
\draw[draw=none,fill=steelblue31119180] (axis cs:5.99104879823391,0) rectangle (axis cs:8.97274903627031,2.2546495418525);
\draw[draw=none,fill=steelblue31119180] (axis cs:8.97274903627031,0) rectangle (axis cs:11.9544492743067,1.69098715638937);
\draw[draw=none,fill=steelblue31119180] (axis cs:11.9544492743067,0) rectangle (axis cs:14.9361495123431,1.97281834912093);
\draw[draw=none,fill=steelblue31119180] (axis cs:14.9361495123431,0) rectangle (axis cs:17.9178497503795,1.40915596365781);
\draw[draw=none,fill=steelblue31119180] (axis cs:17.9178497503795,0) rectangle (axis cs:20.8995499884159,0.845493578194687);
\draw[draw=none,fill=steelblue31119180] (axis cs:20.8995499884159,0) rectangle (axis cs:23.8812502264523,0.281831192731562);
\draw[draw=none,fill=steelblue31119180] (axis cs:23.8812502264523,0) rectangle (axis cs:26.8629504644887,1.40915596365781);
\draw[draw=none,fill=steelblue31119180] (axis cs:26.8629504644887,0) rectangle (axis cs:29.8446507025251,1.69098715638937);
\draw[draw=none,fill=steelblue31119180] (axis cs:29.8446507025251,0) rectangle (axis cs:32.8263509405615,2.53648073458406);
\draw[draw=none,fill=steelblue31119180] (axis cs:32.8263509405615,0) rectangle (axis cs:35.8080511785979,0.281831192731562);
\draw[draw=none,fill=steelblue31119180] (axis cs:35.8080511785979,0) rectangle (axis cs:38.7897514166343,0.845493578194686);
\draw[draw=none,fill=steelblue31119180] (axis cs:38.7897514166343,0) rectangle (axis cs:41.7714516546707,0.563662385463125);
\draw[draw=none,fill=steelblue31119180] (axis cs:41.7714516546707,0) rectangle (axis cs:44.7531518927071,0.845493578194686);
\draw[draw=none,fill=steelblue31119180] (axis cs:44.7531518927071,0) rectangle (axis cs:47.7348521307435,0.563662385463124);
\draw[draw=none,fill=steelblue31119180] (axis cs:47.7348521307435,0) rectangle (axis cs:50.7165523687799,0.563662385463124);
\draw[draw=none,fill=steelblue31119180] (axis cs:50.7165523687799,0) rectangle (axis cs:53.6982526068163,0.845493578194686);
\draw[draw=none,fill=steelblue31119180] (axis cs:53.6982526068163,0) rectangle (axis cs:56.6799528448527,0);
\draw[draw=none,fill=steelblue31119180] (axis cs:56.6799528448527,0) rectangle (axis cs:59.6616530828891,0);
\draw[draw=none,fill=steelblue31119180] (axis cs:59.6616530828891,0) rectangle (axis cs:62.6433533209255,1.12732477092625);
\draw[draw=none,fill=steelblue31119180] (axis cs:62.6433533209255,0) rectangle (axis cs:65.625053558962,0.281831192731563);
\draw[draw=none,fill=steelblue31119180] (axis cs:65.6250535589619,0) rectangle (axis cs:68.6067537969983,0.563662385463122);
\draw[draw=none,fill=steelblue31119180] (axis cs:68.6067537969983,0) rectangle (axis cs:71.5884540350347,1.12732477092625);
\draw[draw=none,fill=steelblue31119180] (axis cs:71.5884540350347,0) rectangle (axis cs:74.5701542730711,0.845493578194688);
\draw[draw=none,fill=steelblue31119180] (axis cs:74.5701542730711,0) rectangle (axis cs:77.5518545111075,0.281831192731561);
\draw[draw=none,fill=steelblue31119180] (axis cs:77.5518545111075,0) rectangle (axis cs:80.5335547491439,0.845493578194688);
\draw[draw=none,fill=steelblue31119180] (axis cs:80.5335547491439,0) rectangle (axis cs:83.5152549871803,2.2546495418525);
\draw[draw=none,fill=steelblue31119180] (axis cs:83.5152549871803,0) rectangle (axis cs:86.4969552252168,0.845493578194684);
\draw[draw=none,fill=steelblue31119180] (axis cs:86.4969552252167,0) rectangle (axis cs:89.4786554632531,0.563662385463125);
\draw (axis cs:45,3.4) node[
scale=1.25,
text=black,
rotate=0.0
]{ $t = 14$ days};
\end{axis}

\end{tikzpicture}\\
	
\begin{tikzpicture}

\definecolor{darkgray176}{RGB}{176,176,176}
\definecolor{steelblue31119180}{RGB}{31,119,180}

\begin{axis}[
tick align=outside,
tick pos=left,
x grid style={darkgray176},
xlabel={Fiber direction [degrees]},
xmin=0, xmax=90,
xtick style={color=black},
xtick={0,15,30,45,60,75,90},
xticklabels={0,15,30,45,60,75,90},
y grid style={darkgray176},
ylabel={Fiber amount [\%]},
ymin=0, ymax=30.0,
ytick style={color=black},
ytick={0.0,5.0,10.0,15.0,20.0,25.0,30.0,35.0},
yticklabels={0.0,5.0,10.0,15.0,20.0,25.0,30.0,35.0}
]
\draw[draw=none,fill=steelblue31119180] (axis cs:0.0195121159831095,0) rectangle (axis cs:3.00165527763844,29.5089790227082);
\draw[draw=none,fill=steelblue31119180] (axis cs:3.00165527763844,0) rectangle (axis cs:5.98379843929377,2.34730514953361);
\draw[draw=none,fill=steelblue31119180] (axis cs:5.98379843929377,0) rectangle (axis cs:8.96594160094911,0);
\draw[draw=none,fill=steelblue31119180] (axis cs:8.96594160094911,0) rectangle (axis cs:11.9480847626044,0.33532930707623);
\draw[draw=none,fill=steelblue31119180] (axis cs:11.9480847626044,0) rectangle (axis cs:14.9302279242598,0);
\draw[draw=none,fill=steelblue31119180] (axis cs:14.9302279242598,0) rectangle (axis cs:17.9123710859151,0);
\draw[draw=none,fill=steelblue31119180] (axis cs:17.9123710859151,0) rectangle (axis cs:20.8945142475704,0);
\draw[draw=none,fill=steelblue31119180] (axis cs:20.8945142475704,0) rectangle (axis cs:23.8766574092258,0);
\draw[draw=none,fill=steelblue31119180] (axis cs:23.8766574092258,0) rectangle (axis cs:26.8588005708811,0);
\draw[draw=none,fill=steelblue31119180] (axis cs:26.8588005708811,0) rectangle (axis cs:29.8409437325364,0);
\draw[draw=none,fill=steelblue31119180] (axis cs:29.8409437325364,0) rectangle (axis cs:32.8230868941918,0);
\draw[draw=none,fill=steelblue31119180] (axis cs:32.8230868941918,0) rectangle (axis cs:35.8052300558471,0);
\draw[draw=none,fill=steelblue31119180] (axis cs:35.8052300558471,0) rectangle (axis cs:38.7873732175024,0.335329307076229);
\draw[draw=none,fill=steelblue31119180] (axis cs:38.7873732175024,0) rectangle (axis cs:41.7695163791578,0);
\draw[draw=none,fill=steelblue31119180] (axis cs:41.7695163791578,0) rectangle (axis cs:44.7516595408131,0);
\draw[draw=none,fill=steelblue31119180] (axis cs:44.7516595408131,0) rectangle (axis cs:47.7338027024684,0);
\draw[draw=none,fill=steelblue31119180] (axis cs:47.7338027024684,0) rectangle (axis cs:50.7159458641238,0);
\draw[draw=none,fill=steelblue31119180] (axis cs:50.7159458641238,0) rectangle (axis cs:53.6980890257791,0);
\draw[draw=none,fill=steelblue31119180] (axis cs:53.6980890257791,0) rectangle (axis cs:56.6802321874344,0);
\draw[draw=none,fill=steelblue31119180] (axis cs:56.6802321874344,0) rectangle (axis cs:59.6623753490898,0);
\draw[draw=none,fill=steelblue31119180] (axis cs:59.6623753490898,0) rectangle (axis cs:62.6445185107451,0);
\draw[draw=none,fill=steelblue31119180] (axis cs:62.6445185107451,0) rectangle (axis cs:65.6266616724004,0);
\draw[draw=none,fill=steelblue31119180] (axis cs:65.6266616724004,0) rectangle (axis cs:68.6088048340558,0);
\draw[draw=none,fill=steelblue31119180] (axis cs:68.6088048340558,0) rectangle (axis cs:71.5909479957111,0);
\draw[draw=none,fill=steelblue31119180] (axis cs:71.5909479957111,0) rectangle (axis cs:74.5730911573664,0);
\draw[draw=none,fill=steelblue31119180] (axis cs:74.5730911573664,0) rectangle (axis cs:77.5552343190217,0);
\draw[draw=none,fill=steelblue31119180] (axis cs:77.5552343190218,0) rectangle (axis cs:80.5373774806771,0);
\draw[draw=none,fill=steelblue31119180] (axis cs:80.5373774806771,0) rectangle (axis cs:83.5195206423324,0);
\draw[draw=none,fill=steelblue31119180] (axis cs:83.5195206423324,0) rectangle (axis cs:86.5016638039878,0);
\draw[draw=none,fill=steelblue31119180] (axis cs:86.5016638039878,0) rectangle (axis cs:89.4838069656431,1.00598792122869);
\draw (axis cs:45,25.5) node[
scale=1.25,
text=black,
rotate=0.0
]{ $t = 28$ days};
\end{axis}

\end{tikzpicture}
	  \
\begin{tikzpicture}

\definecolor{darkgray176}{RGB}{176,176,176}
\definecolor{steelblue31119180}{RGB}{31,119,180}

\begin{axis}[
tick align=outside,
tick pos=left,
x grid style={darkgray176},
xlabel={Fiber direction [degrees]},
xmin=0, xmax=90,
xtick style={color=black},
xtick={0,15,30,45,60,75,90},
xticklabels={0,15,30,45,60,75,90},
y grid style={darkgray176},
ylabel={Fiber amount [\%]},
ymin=0, ymax=6.0,
ytick style={color=black},
ytick={0.0,1.0,2.0,3.0,4.0,5.0,6.0},
yticklabels={0.0,1.0,2.0,3.0,4.0,5.0,6.0}
]
\draw[draw=none,fill=steelblue31119180] (axis cs:0.419325445566406,0) rectangle (axis cs:3.35145757811392,5.41347222958367);
\draw[draw=none,fill=steelblue31119180] (axis cs:3.35145757811392,0) rectangle (axis cs:6.28358971066142,2.70673611479183);
\draw[draw=none,fill=steelblue31119180] (axis cs:6.28358971066143,0) rectangle (axis cs:9.21572184320894,2.70673611479183);
\draw[draw=none,fill=steelblue31119180] (axis cs:9.21572184320893,0) rectangle (axis cs:12.1478539757564,2.70673611479183);
\draw[draw=none,fill=steelblue31119180] (axis cs:12.1478539757564,0) rectangle (axis cs:15.079986108304,1.89471528035428);
\draw[draw=none,fill=steelblue31119180] (axis cs:15.0799861083039,0) rectangle (axis cs:18.0121182408515,1.08269444591673);
\draw[draw=none,fill=steelblue31119180] (axis cs:18.0121182408515,0) rectangle (axis cs:20.944250373399,0.81202083443755);
\draw[draw=none,fill=steelblue31119180] (axis cs:20.944250373399,0) rectangle (axis cs:23.8763825059465,1.35336805739592);
\draw[draw=none,fill=steelblue31119180] (axis cs:23.8763825059465,0) rectangle (axis cs:26.808514638494,1.35336805739592);
\draw[draw=none,fill=steelblue31119180] (axis cs:26.808514638494,0) rectangle (axis cs:29.7406467710415,0.81202083443755);
\draw[draw=none,fill=steelblue31119180] (axis cs:29.7406467710415,0) rectangle (axis cs:32.672778903589,0.541347222958367);
\draw[draw=none,fill=steelblue31119180] (axis cs:32.672778903589,0) rectangle (axis cs:35.6049110361365,0.270673611479183);
\draw[draw=none,fill=steelblue31119180] (axis cs:35.6049110361365,0) rectangle (axis cs:38.537043168684,0.81202083443755);
\draw[draw=none,fill=steelblue31119180] (axis cs:38.537043168684,0) rectangle (axis cs:41.4691753012315,0.541347222958367);
\draw[draw=none,fill=steelblue31119180] (axis cs:41.4691753012315,0) rectangle (axis cs:44.401307433779,0.270673611479183);
\draw[draw=none,fill=steelblue31119180] (axis cs:44.401307433779,0) rectangle (axis cs:47.3334395663266,0);
\draw[draw=none,fill=steelblue31119180] (axis cs:47.3334395663266,0) rectangle (axis cs:50.2655716988741,0.541347222958367);
\draw[draw=none,fill=steelblue31119180] (axis cs:50.2655716988741,0) rectangle (axis cs:53.1977038314216,1.08269444591673);
\draw[draw=none,fill=steelblue31119180] (axis cs:53.1977038314216,0) rectangle (axis cs:56.1298359639691,1.08269444591673);
\draw[draw=none,fill=steelblue31119180] (axis cs:56.1298359639691,0) rectangle (axis cs:59.0619680965166,0.81202083443755);
\draw[draw=none,fill=steelblue31119180] (axis cs:59.0619680965166,0) rectangle (axis cs:61.9941002290641,0.541347222958365);
\draw[draw=none,fill=steelblue31119180] (axis cs:61.9941002290641,0) rectangle (axis cs:64.9262323616116,1.08269444591674);
\draw[draw=none,fill=steelblue31119180] (axis cs:64.9262323616116,0) rectangle (axis cs:67.8583644941591,0.81202083443755);
\draw[draw=none,fill=steelblue31119180] (axis cs:67.8583644941591,0) rectangle (axis cs:70.7904966267066,0);
\draw[draw=none,fill=steelblue31119180] (axis cs:70.7904966267066,0) rectangle (axis cs:73.7226287592541,0.81202083443755);
\draw[draw=none,fill=steelblue31119180] (axis cs:73.7226287592541,0) rectangle (axis cs:76.6547608918016,0.541347222958367);
\draw[draw=none,fill=steelblue31119180] (axis cs:76.6547608918016,0) rectangle (axis cs:79.5868930243492,0.81202083443755);
\draw[draw=none,fill=steelblue31119180] (axis cs:79.5868930243492,0) rectangle (axis cs:82.5190251568967,1.6240416688751);
\draw[draw=none,fill=steelblue31119180] (axis cs:82.5190251568967,0) rectangle (axis cs:85.4511572894442,0.541347222958367);
\draw[draw=none,fill=steelblue31119180] (axis cs:85.4511572894442,0) rectangle (axis cs:88.3832894219917,0.541347222958367);
\draw (axis cs:45,5.1) node[
scale=1.25,
text=black,
rotate=0.0
]{ $t = 28$ days};
\end{axis}

\end{tikzpicture}
         \end{tabular}}     
        \caption{(a) Schematic highlighting regions where collagen orientations are quantified. (b) Collagen fiber orientation at various time points. Figures on the left side fiber distribution at the middle of the specimen, while on the right side is the distribution at the leg region.}
	\label{fig:collagen_orientation}    
\end{figure}
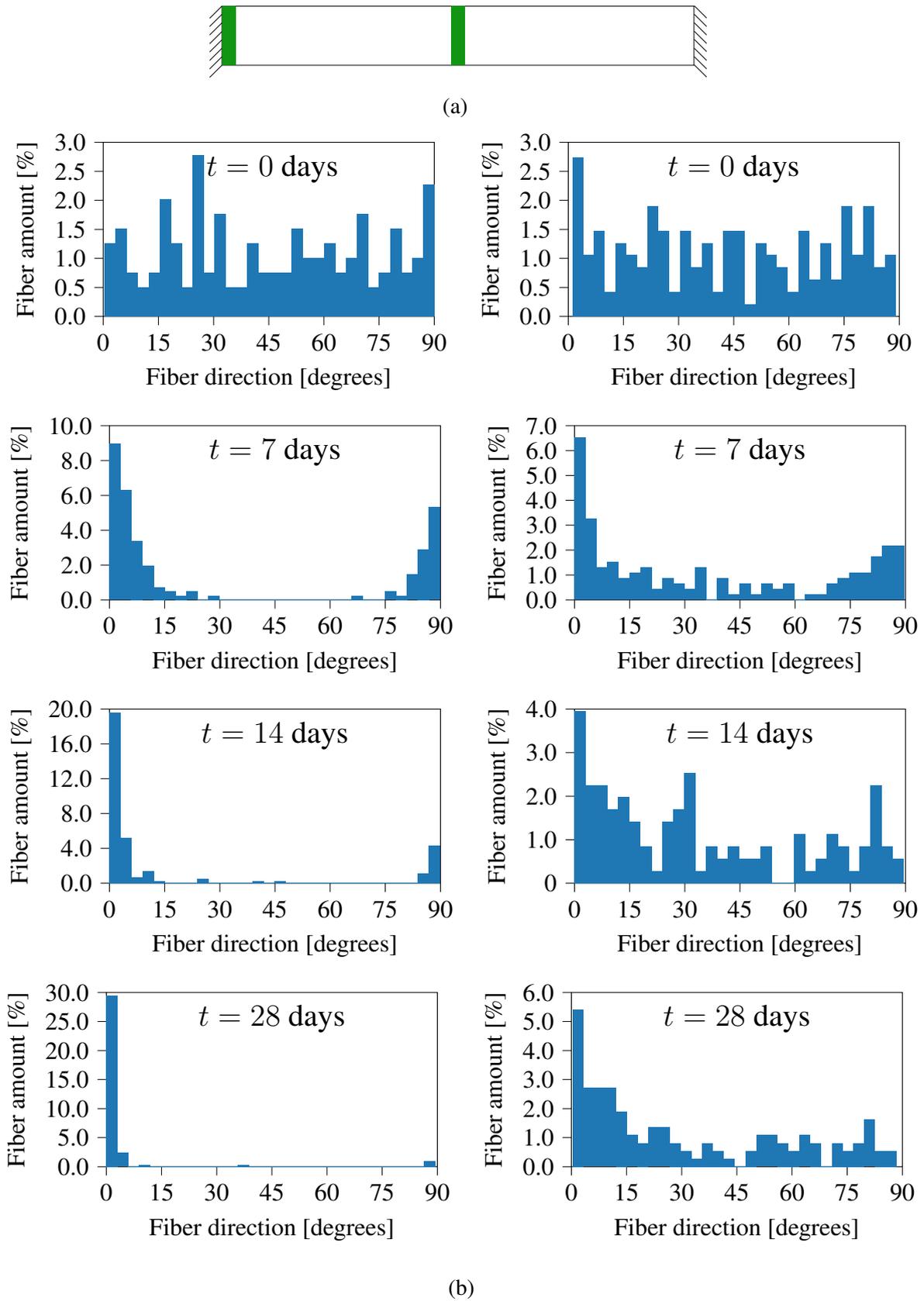

\subsection{Biaxially constrained tissue implant}
 
In the second example, we examine more complex mechanical loading conditions. Here, we investigate a geometry with biaxial constraints. This setup was used to experimentally investigate tensional homeostasis \cite{Eichinger_2020}. Later, Holthusen et al.\ \cite{Holthusen_2023} framework demonstrated an excellent capability in replicating experimental results. In this work, we go one step further by investigating this boundary value problem in the context of tissue maturation over a period of 28 days. Our goal is to examine the ability of the model proposed here to compute a problem with complex loading conditions. 

\begin{figure}[ht]
	\centering
	\includegraphics[scale=0.75]{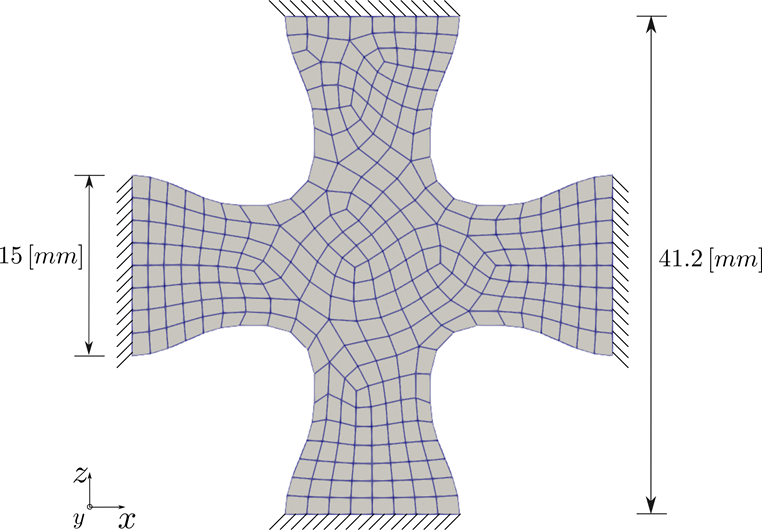}
	\caption{Schematic representation of the boundary value problem for a biaxially constrained tissue construct.}
	\label{fig:biaxial_BVP}
\end{figure}

The boundary value problem that we investigate here is illustrated in Fig.\ \ref{fig:biaxial_BVP}. The figure illustrates the biaxial boundary conditions applied to the specimen. We compute the maturation for a period of 28 days. At the time point of $t = 17$ days,  a biaxial load perturbation of $20 \%$ is applied. The geometry of the specimen is discretized using 626 cubic elements, with two elements along the thickness (y-direction).  

The next step is to define the values of the material parameters. The parameters can be divided into two groups: i) stiffness-like parameters, and ii) parameters describing the evolution equations. For the first group, namely $\lambda$, $\mu$, $ k_1 $, and $ k_2 $ we used the values identified in \cite{Holthusen_2023}. Furthermore, for the parameters $\sigma_{\mathrm{g, 0}}$ and $v_{\mathrm{g}}$ we applied the values identified in \cite{Holthusen_2023}. For the remaining parameters, we have chosen values that give us a reasonable physiological response. The parameter values and the corresponding references are listed in Table \ref{table:5_2}. 

\begin{table}[H]
	\centering \renewcommand\cellalign{lc}
	\setlength{\tabcolsep}{10 pt}
	\begin{tabular}{l l l l l}
		\toprule
		Symbol & Value & Units & Reference \\
		\midrule
		$ \lambda $ & $818$ & $\mathrm{[\SI{}{\micro N} / mm^{2}]}$ & \cite{Holthusen_2023}  \\
		$ \mu  $ & $982$ & $\mathrm{[\SI{}{\micro N} / mm^{2}]}$ & \cite{Holthusen_2023}  \\
		$ k_1 $  & $3351$ & $\mathrm{[\SI{}{\micro N} / mm^{2}]}$ & \cite{Holthusen_2023}  \\
		$ k_2 $ & $14996$ & $[-]$ & \cite{Holthusen_2023}   \\
		$ \kappa $ & $0.10$ & $[-]$ & Selected  \\
		$\sigma_{\mathrm{g, 0}}$  & $22.9 $ & $\mathrm{[\SI{}{\micro N} / mm^{2}]}$ & \cite{Holthusen_2023} \\
		$r_{\mathrm{1}}$ &  $10$ & $\mathrm{[\SI{}{\micro N} / mm^{2}]}$ & Selected \\
		$ a_{1} $ & $2 \times 10^{-3}$ & $\mathrm{[\SI{}{\micro\gram} / cells]}$ &  Selected \\
		$ \tau $  & $7$ & $\mathrm{[days]}$ & Selected \\
		$ h  $ & $1.65$ & $[-]$ & Selected \\
		$ a_{2} $ & $5 \times 10^{-6}$ & $\mathrm{[mm^{3}/cells/day]}$ & Selected \\
		$ {\psi}_{\mathrm{crit}} $ & $3 \times 10^{-5}$ & $\mathrm{[J/\SI{}{\micro\gram}]}$ & Selected \\
		$ {\rho}_{\mathrm{th}} $ & $10$ & $\mathrm{[\SI{}{\micro\gram} / mm^{3}]}$ & Selected \\
		$ {\rho}_{\mathrm{co, f}} $ & $38.7$ & $\mathrm{[\SI{}{\micro\gram} / mm^{3}]}$ & Selected \\
		$ c_{\mathrm{cell}}  $ & $15 \times 10^{3}$ & $\mathrm{[cells/mm^{3}]}$ & \cite{Hermans_etal_2022} \\
		$\beta_{\mathrm{g}}$  & $1 $ & $\mathrm{[\SI{}{\micro N} / mm^{2}]}$  & Selected \\
		$\eta_{\mathrm{g}}$  & $100 $ & $\mathrm{[days]}$  & Selected \\
		$\eta_{\mathrm{s}}$  & $5 $ & $\mathrm{[days]}$  & Selected \\
		$v_{\mathrm{g}}$ &  $1 $ & $[-]$  & \cite{Holthusen_2023} \\
		\bottomrule
	\end{tabular}
	\caption{Material parameters for modeling the maturation of biaxially constrained tissue stripe.}
	\label{table:5_2}
\end{table}

\begin{figure}[H]
	\centering
	\captionsetup[subfigure]{labelformat=empty}
	\subfloat[ $0$  [days{]}]{\includegraphics[scale=0.55]{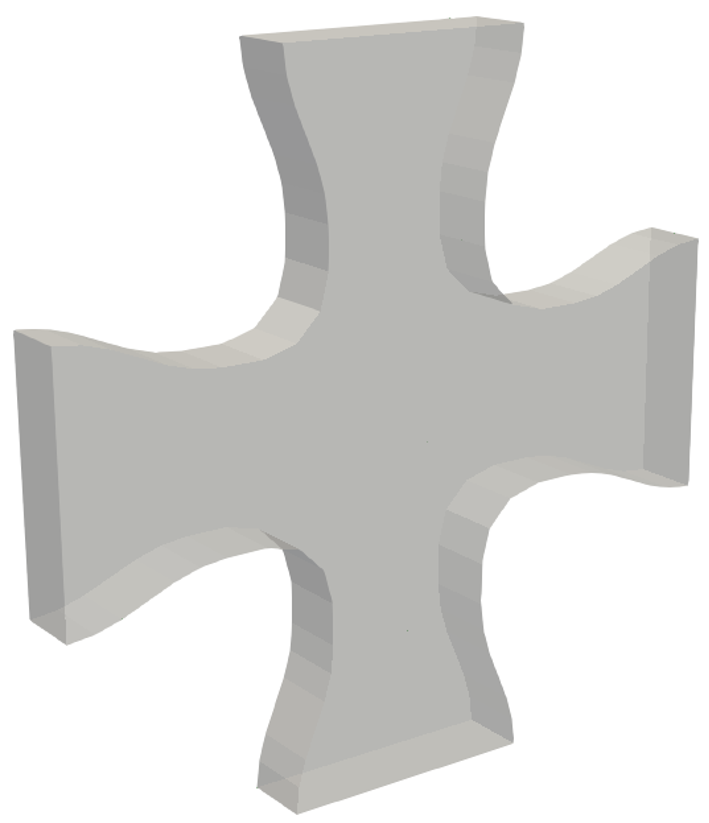}} \qquad	\qquad
	\subfloat[ $5$  [days{]}]{\includegraphics[scale=0.55]{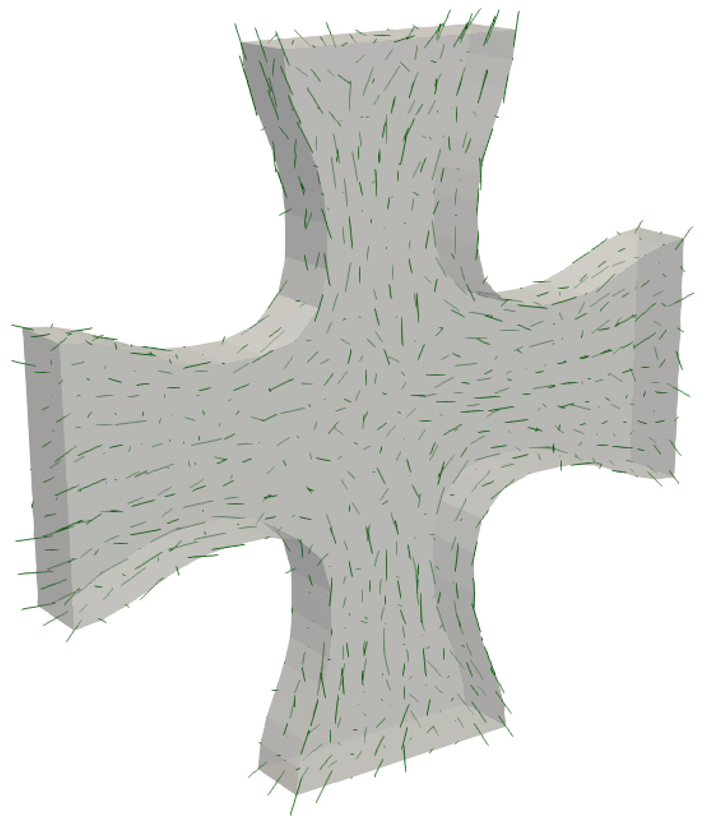}}
	\vspace{48pt}
	
	\subfloat[$17^{-}$   [days{]}]{\includegraphics[scale=0.55]{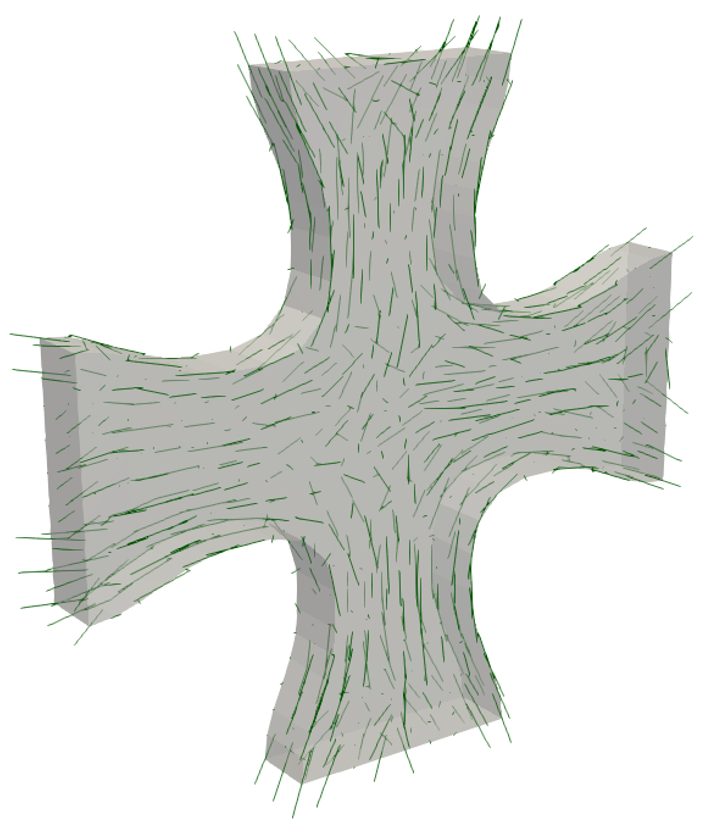}}	\qquad \qquad
	\subfloat[$28$ [days{]}]{\includegraphics[scale=0.55]{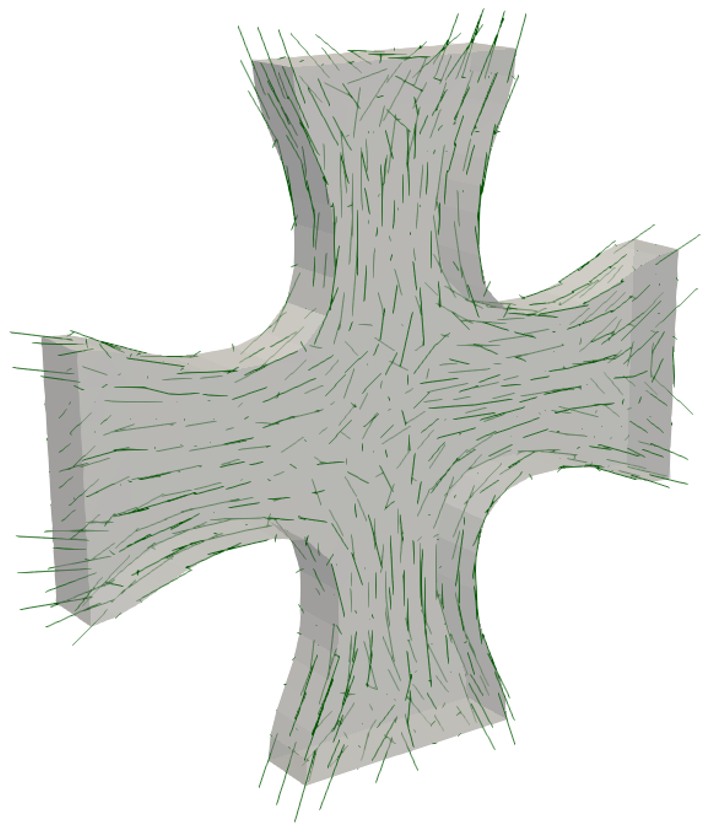}} 
	
	\caption{Evolution of collagen fiber density and orientation during the maturation process of a biaxially constrained tissue.}    
	\label{fig:collagen_evolution_biaxial}    
\end{figure}

\begin{figure}[H]
	\centering
	\captionsetup[subfigure]{labelformat=empty}
	\subfloat[ $5$  [days{]}]{\includegraphics[scale=0.55]{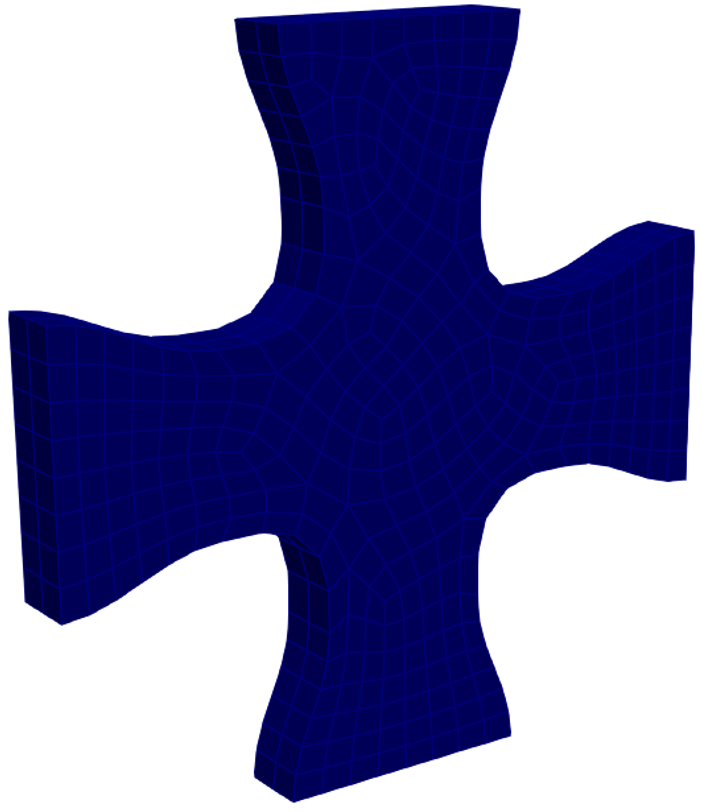}} \qquad	\qquad
	\subfloat[ $10$  [days{]}]{\includegraphics[scale=0.55]{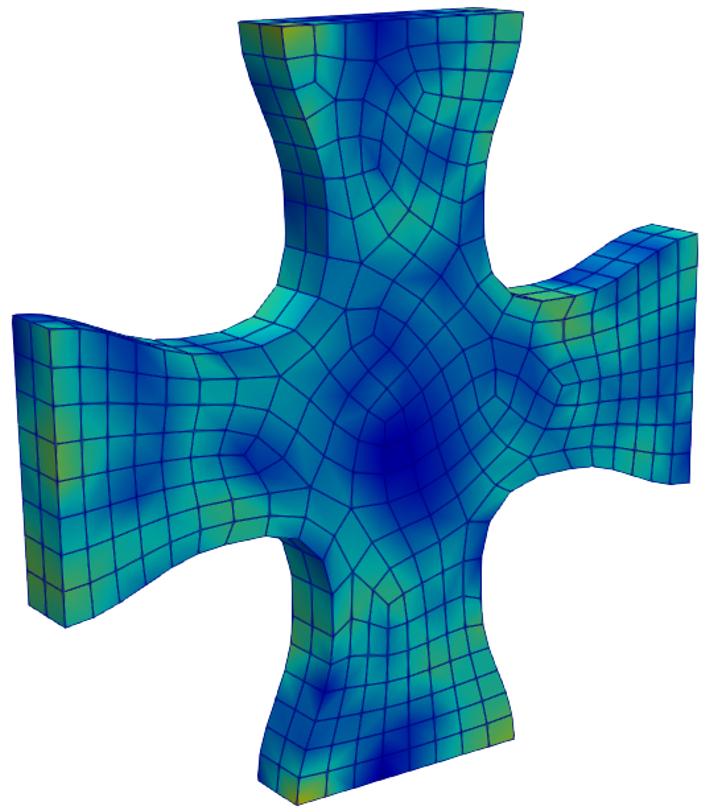}}	
	\vspace{36pt}
	
	\subfloat[ $17^{-}$  [days{]}]{\includegraphics[scale=0.55]{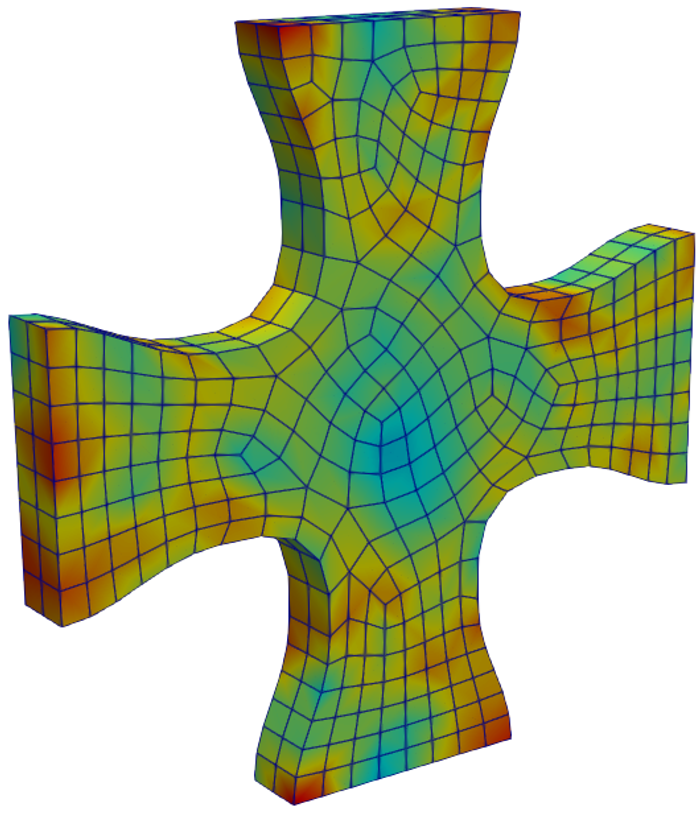}}	\qquad \qquad
	\subfloat[$28$  [days{]}]{\includegraphics[scale=0.55]{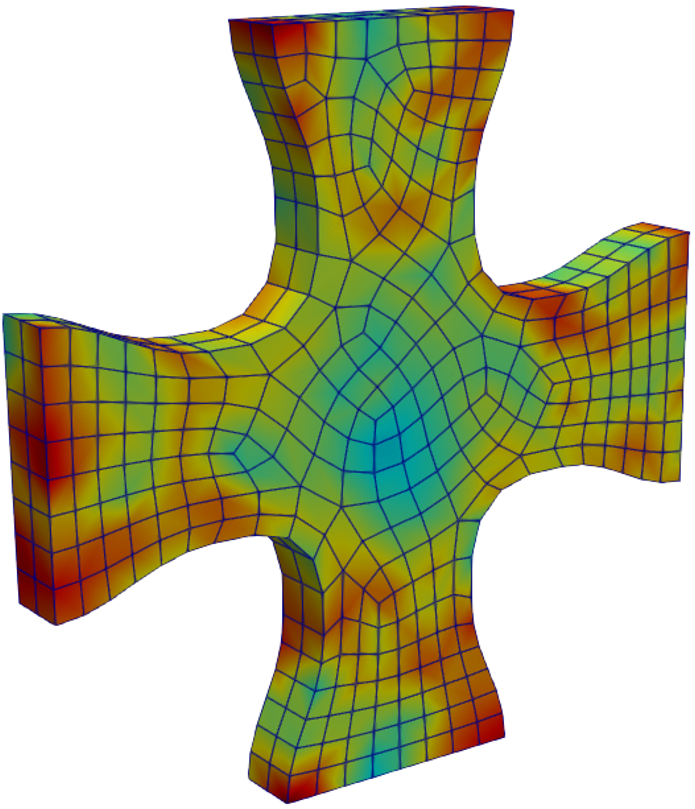}} 
	\vspace{12pt}
	
	\subfloat[]{\includegraphics[scale=0.6]{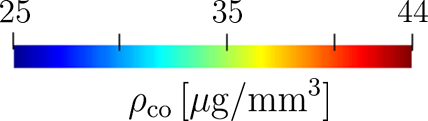}}
	
	\caption{Evolution of collagen density during the maturation process of biaxially constrained tissue.}
	\label{fig:collagen_density_biaxial}    
\end{figure}

Fig.\ \ref{fig:collagen_evolution_biaxial} shows the evolution of collagen fibers. Similar to the first example, collagen density is represented using the length of the green lines. The collagen density distribution at these time points is shown in the contour plots in Fig.\ \ref{fig:collagen_density_biaxial}. In this example, we observe that fibers are highly oriented in the axial section, while in the middle region where there is biaxial loading conditions, collagen fibers have a dispersed orientation. Collagen orientations resemble the results in \cite{Holthusen_2023}. A closer look at the collagen density distribution in Fig.\ \ref{fig:collagen_evolution_biaxial}, shows lower collagen density in the middle region. This can be explained by Eq.\ \ref{eq:rho_mech} where mechanical stimulation of collagen growth is defined as a function of the energy term ${\psi}_{\mathrm{co, m}}$. Fiber dispersion lowers the value of the strain energy ${\psi}_{\mathrm{co, m}}$.

\begin{figure}[H]
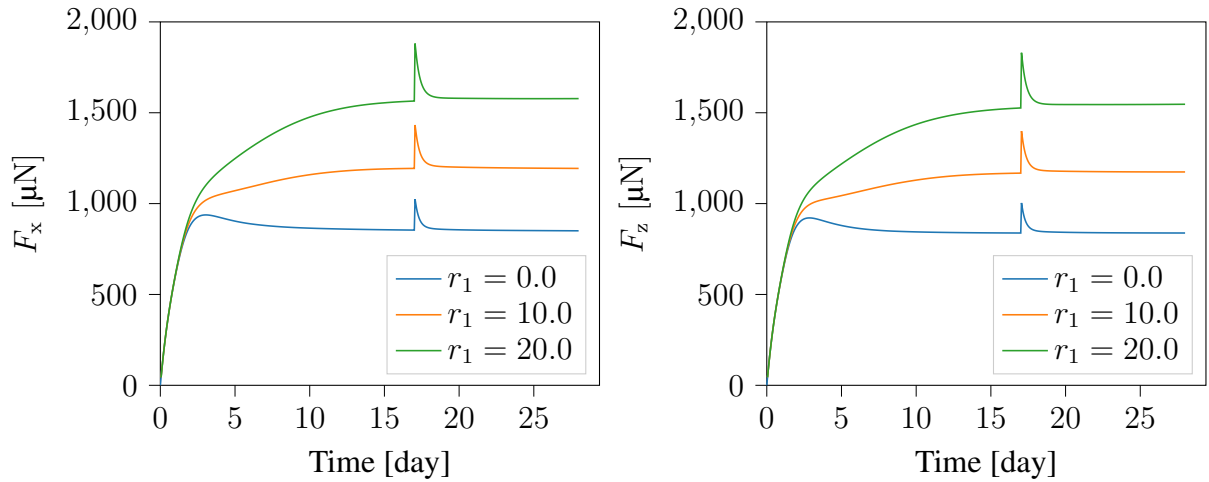

	\centering
	\subfloat{\input{figures/reaction_force_Fx.tex}}
	\subfloat{\input{figures/reaction_force_Fz.tex}}
	
	\caption{The evolution of reaction forces along the sample boundaries in the (left) x-direction, and (right) z-direction.}
	\label{fig:biaxial_force_response}    
\end{figure}

\begin{figure}[H]
	\centering
	\captionsetup[subfigure]{labelformat=empty}
	\subfloat[$5$ [days{]}]{\includegraphics[scale=0.55]{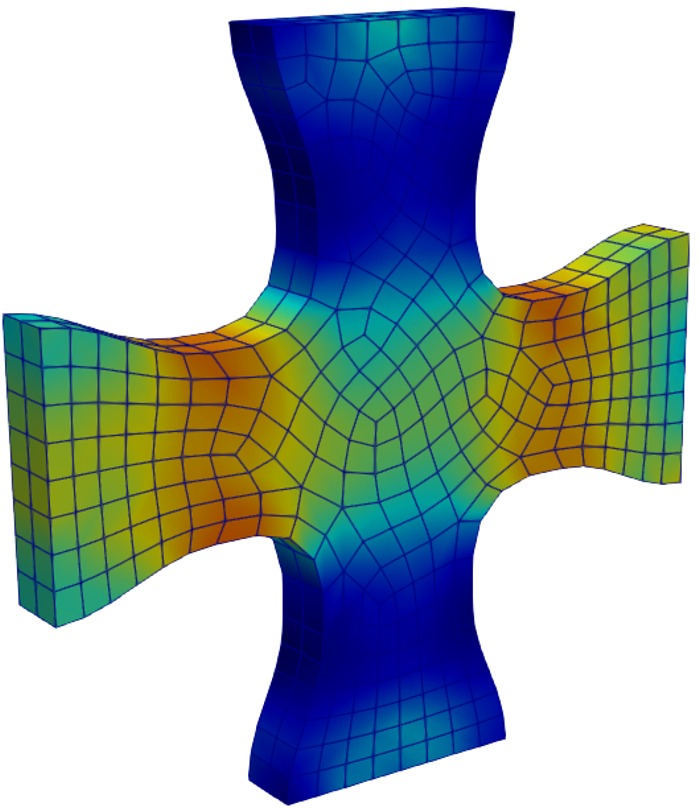}} \qquad	\qquad
	\subfloat[$17^{-}$  [days{]}]{\includegraphics[scale=0.55]{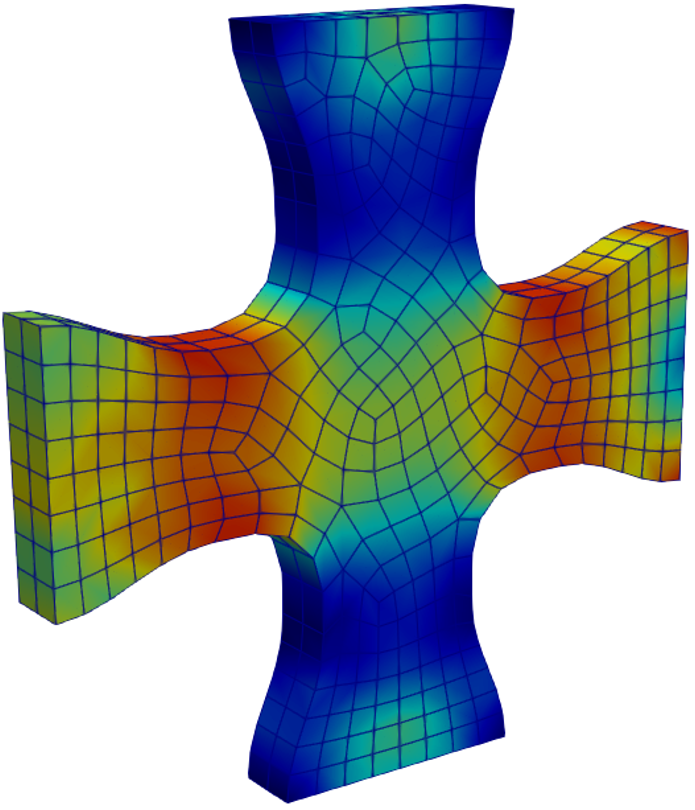}}	
	\vspace{36pt}
	
	\subfloat[$17^{+}$  [days{]}]{\includegraphics[scale=0.55]{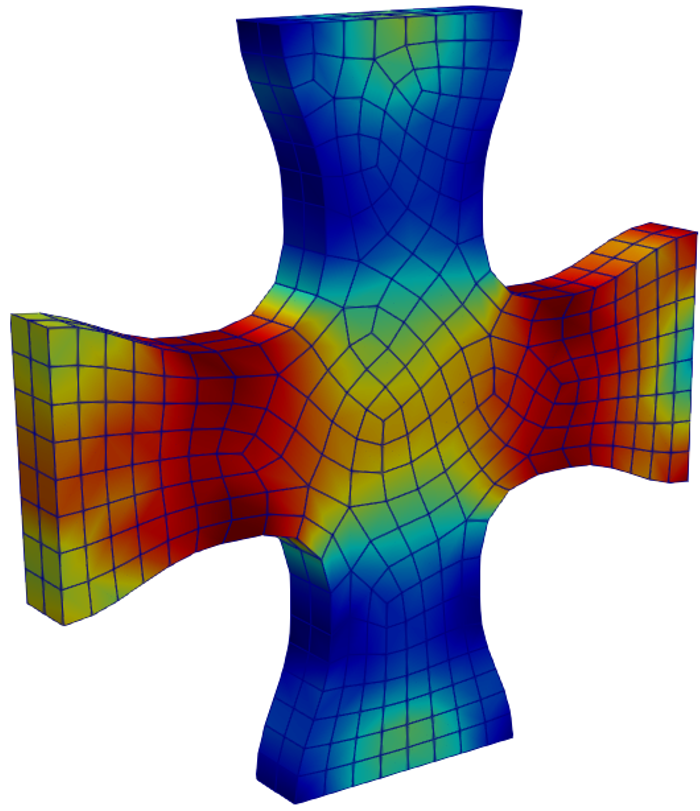}}	\qquad \qquad
	\subfloat[$28$ [days{]}]{\includegraphics[scale=0.55]{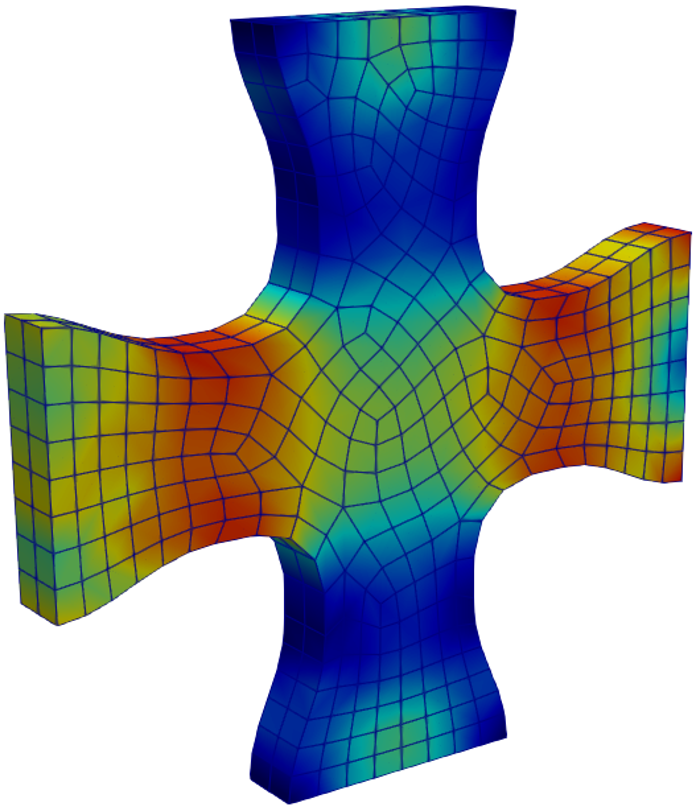}} 
	\vspace{12pt}
	
	\subfloat[]{\includegraphics[scale=0.5]{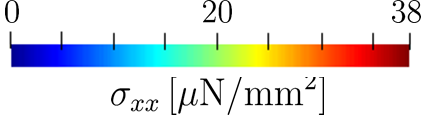}}
	
	\caption{Evolution of the Cauchy stress in the $x$ direction at different time steps. $17^{-}$ refers to the last time step before the $20 \%$ load perturbation, and $17^{+}$ refers to first time step after the load perturbation.}
	\label{fig:sigma_xx_biaxial}    
\end{figure}

\begin{figure}[H]
	\centering
	\captionsetup[subfigure]{labelformat=empty}
	\subfloat[$5$ [days{]}]{\includegraphics[scale=0.55]{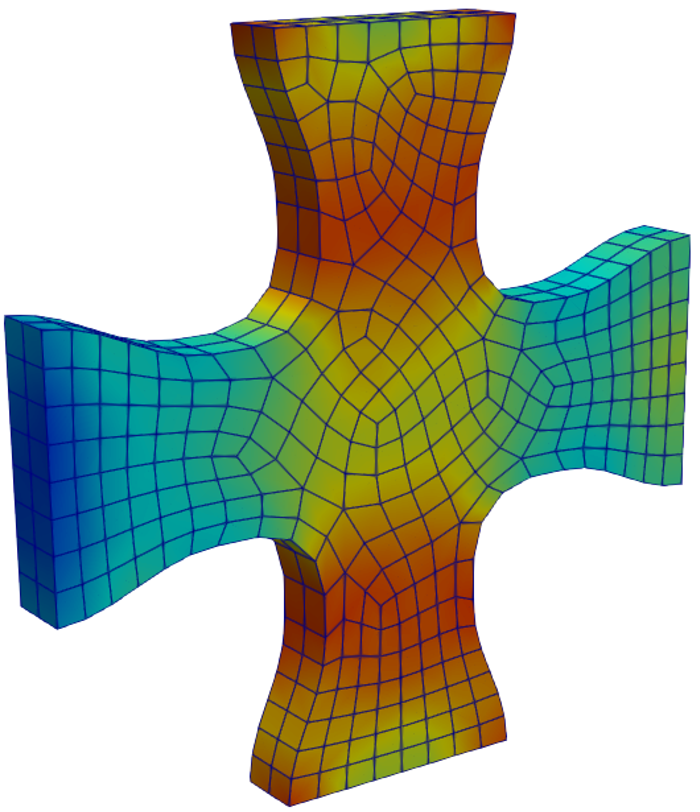}} \qquad	\qquad
	\subfloat[$17^{-}$  [days{]}]{\includegraphics[scale=0.55]{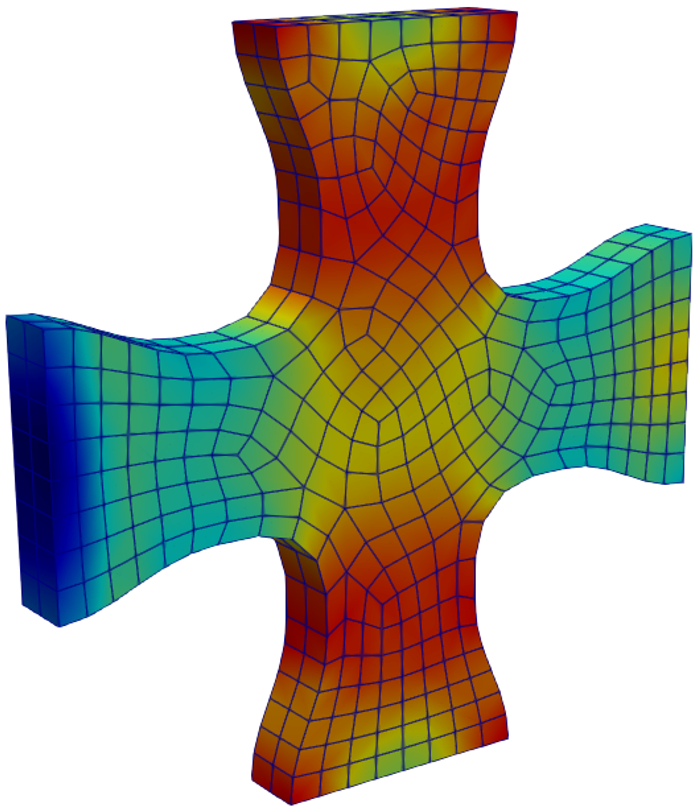}}	
	\vspace{36pt}
	
	\subfloat[$17^{+}$  [days{]}]{\includegraphics[scale=0.55]{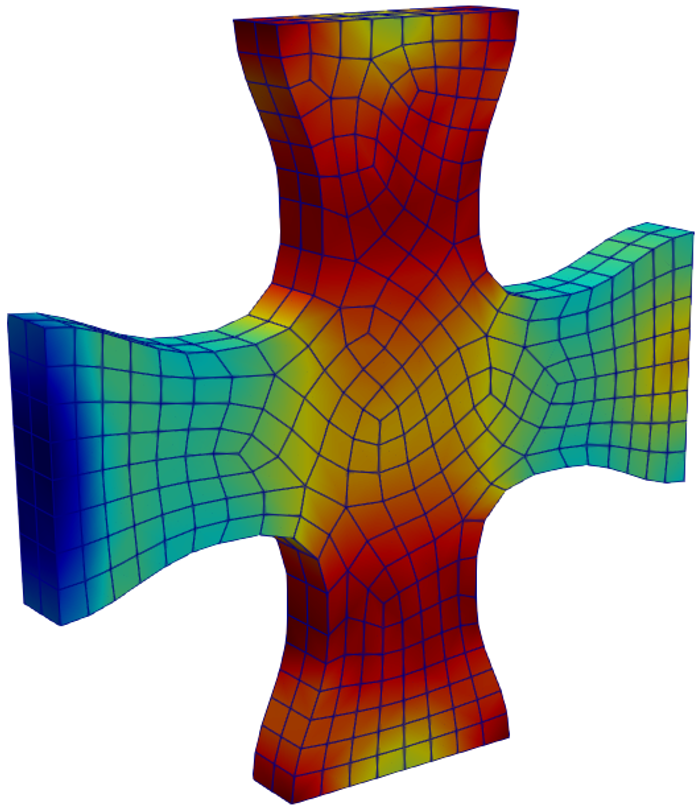}}	\qquad \qquad
	\subfloat[$28$ [days{]}]{\includegraphics[scale=0.55]{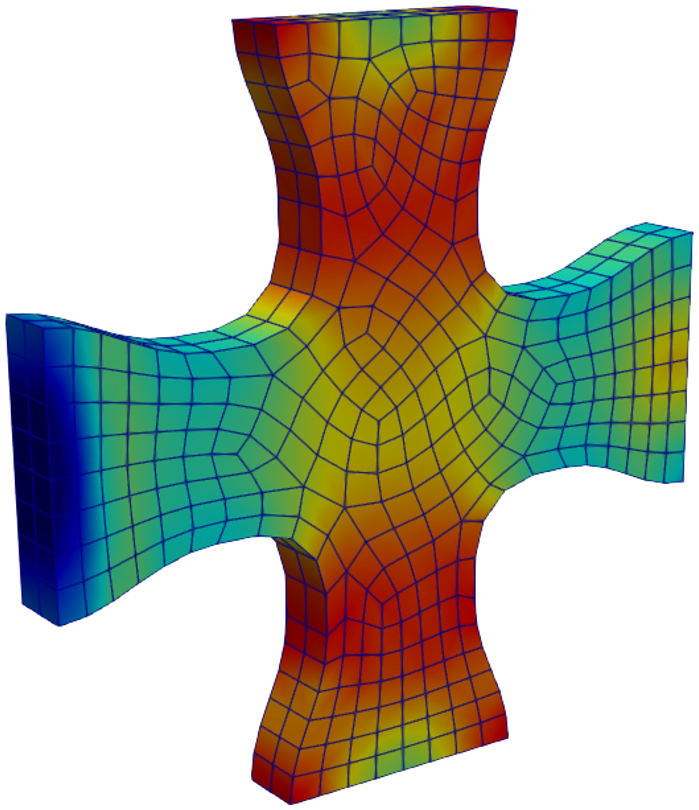}} 
	\vspace{12pt}
	
	\subfloat{\includegraphics[scale=0.5]{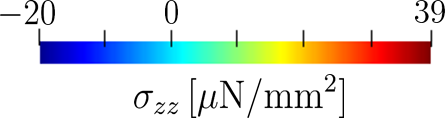}}
	
	\caption{Evolution of the Cauchy stress in the $z$ direction at different time steps. $17^{-}$ refers to the last time step before the $20 \%$ load perturbation, and $17^{+}$ refers to first time step after the load perturbation.}
	\label{fig:sigma_zz_biaxial}    
\end{figure}

The stress contours in Fig.\ \ref{fig:sigma_xx_biaxial} and \ref{fig:sigma_zz_biaxial} show the evolution of the stresses along the X and Z-directions respectively. In both figures, we observe a significant increase in the stress following the load perturbation. However, as material returns to the homeostatic stress levels, we observe that stress at time points $17^{-}$ and $28$ days are almost identical. $17^{-}$ here refers to the last time point before the load perturbation at $t = 17$. This can be observed in Fig.\ \ref{fig:biaxial_force_response}, where the force response returns to the homeostatic level following the load perturbation.

\pgfplotsset{%
	width=0.46\textwidth,
	height=0.40\textwidth
}
Furthermore, we perform a parameter study on the coupling parameter $r_{\mathrm{1}}$ which is introduced in Eq.\ \ref{homeostatic_stress}. A value of $r_{\mathrm{1}} = 0$ means that the collagen density does not influence the homeostatic stress. The force response plots in Fig.\ \ref{fig:biaxial_force_response} show that the value of $r_{\mathrm{1}}$ significantly influences the force response curves. A higher value leads to a steeper increase in reaction force, in particular during the initial period of the maturation process where we witness a significant increase in collagen density.


\section{Conclusion and Outlook}\label{sec:5}

This work presents a novel framework for modeling growth and remodeling during the maturation of tissue-engineered implants. A key feature is the explicit treatment of collagen, the main structural constituent in soft biological tissues, whose density and orientation evolve over time. Additionally, volumetric growth is governed by a homeostatic surface approach that ensures compliance with the second law of thermodynamics, thereby providing a more comprehensive and physically consistent model compared to earlier efforts \cite{Lamm_2022, Sesa_2023_CBM, Holthusen_2023}. A central theoretical advancement is the dependence of homeostatic stress on local collagen content, allowing for spatial variability in tissue composition. Implementing a single homeostatic surface, combined with two growth potentials defined in a non-associative manner, also affords greater flexibility in handling multiple tissue constituents without resorting to multi-surface formulations.

The model was tested in two numerical examples. In the first, the in vitro maturation of a uniaxially constrained tissue stripe provided experimental data for parameter identification, verifying the model’s ability to capture collagen density evolution and reorientation with reasonable accuracy. Parameters describing overall tissue stiffness were taken from prior work \cite{Sesa_2023_CBM}. In the second example, biaxially constrained tissue was subjected to load perturbation, underscoring the model’s capability to handle more complex boundary conditions.

Looking ahead, the proposed modeling framework can be applied to three-dimensional tissue-engineered implants like vascular grafts and heart valves, offering opportunities for more rigorous validation and parameter tuning under clinically relevant conditions. Expanding the model to reinforced biohybrid implants \cite{Boehm_2023} could further illuminate the mechanobiological effects of different scaffold types. Ultimately, embedding this constitutive approach into a multi-physics fluid-solid-growth framework \cite{Figueroa_2009} would enable even more realistic simulations. Another promising direction involves leveraging automatic model discovery techniques \cite{Holthusen_2024a, Holthusen_2024b} to refine or generalize the growth and remodeling laws, potentially leading to new insights and clinically relevant predictions.

\textbf{Declaration of competing interest}

The authors declare that they have no known competing financial interests or personal relationships that could have appeared to influence the work reported in this paper.

\textbf{Acknowledgment}

 Stefanie Reese and Stefan Jockenhövel gratefully acknowledge the financial support provided by the German Research Foundation (DFG) for project 403471716 "Experimental investigations and modeling of biohybrid heart valves including tissue maturation – from in vitro to in situ tissue engineering" which is part of DFG PAK-961 consortium "Towards a model based control of biohybrid implant maturation". Furthermore, Kevin Linka and Stefanie Reese acknowledge the financial support granted by the DFG for project 465213526 "In-stent restenosis in coronary arteries – computational and data-driven investigations towards translational modeling." In addition, Kevin Linka acknowledges the Emmy Noether Grant 533187597 "Computational Soft Material Mechanics Intelligence" from the DFG.

\bibliographystyle{IEEEtran.bst}
\bibliography{literature}

\end{document}